\documentclass[]{aastex631}

\usepackage{multirow}
\usepackage{amsmath}
\usepackage{booktabs}

\newcommand{\kms}{\ifmmode\,{\rm km}\,{\rm s}^{-1}\else km$\,$s$^{-1}$\fi}
\newcommand{\Rd}{\ifmmode\,R_{\rm d}\else $R_{\rm d}$\fi}
\newcommand{\be}{\begin{equation}}
\newcommand{\ee}{\end{equation}}
\newcommand\ltsima{$\; \buildrel < \over \sim \;$}
\newcommand\ltsim{\lower.5ex\hbox{\ltsima}}
\newcommand\gtsima{$\; \buildrel > \over \sim \;$}
\newcommand\gtsim{\lower.5ex\hbox{\gtsima}}

\newcommand{\magss}{\ifmmode {{{{\rm mag}~{\rm arcsec}}^{-2}}}
             \else {{{mag}$~${arcsec}$^{-2}$}}
             \fi}

\newcommand{\mycomment}[1]{}
\def\sec#1{Sec.~\ref{sec:#1}}
\def\app#1{App.~\ref{sec:#1}}
\def\Fig#1{Fig.~\ref{fig:#1}}
\def\Table#1{Table~\ref{tab:#1}}
\def\Eq#1{Eq.~(\ref{eq:#1})}

\def \HI{\ion{H}{I}\thinspace}
\def \HI {H\,{\sc i}\thinspace}

\def \littleprime{\ifmmode{\scriptscriptstyle \prime }
     \else{\hbox{$\scriptscriptstyle \prime$ }}\fi}
\DeclareMathOperator{\sech}{sech}
\defcitealias{Haynes1984}{HG84}

\begin{document}

\title{The Intrinsic Flattening of Extragalactic Stellar Disks}

\shorttitle{The Intrinsic Flattening of Extragalactic Stellar Disks}
\shortauthors{Favaro et al.}

\correspondingauthor{Jeremy Favaro}
\email{jeremy.favaro@queensu.ca}

\author[0009-0005-6999-2073]{Jeremy Favaro}
\affiliation{Department of Physics, Engineering Physics \& Astronomy, Queen's University, Kingston, ON K7L 3N6, Canada}
\author[0000-0002-8597-6277]{Ste\'phane Courteau}
\affiliation{Department of Physics, Engineering Physics \& Astronomy, Queen's University, Kingston, ON K7L 3N6, Canada}
\author[0000-0002-7398-4907]{S\'ebastien Comer\'on}
\affiliation{Departamento de Astrofísica, Universidad de La Laguna, 38200 La Laguna, Tenerife, Spain}
\affiliation{Instituto de Astrofísica de Canarias, 38205 La Laguna, Tenerife, Spain}
\author[0000-0002-9086-6398]{Connor Stone}
\affiliation{Department of Physics, Universit{\'e} de Montr{\'e}al, Montr{\'e}al, Qu{\' e}bec, Canada}
\affiliation{Mila - Qu{\' e}bec Artificial Intelligence Institute, Montr{\'e}al, Qu{\' e}bec, Canada}
\affiliation{Ciela - Montr{\'e}al Institute for Astrophysical Data Analysis and Machine Learning, Montr{\'e}al, Qu{\' e}bec, Canada}

\begin{abstract}

Highly inclined (edge-on) disk galaxies offer the unique perspective to constrain their intrinsic flattening, $c/a$, where $c$ and $a$ are respectively the vertical and long radial axes of the disk measured at suitable stellar densities. 
The ratio $c/a$ is a necessary quantity in the assessment of galaxy inclinations, three-dimensional structural reconstructions, total masses, as well as a constraint to galaxy formation models.  
$3.6 \ \mu \text{m}$ maps of 133 edge-on spiral galaxies from the Spitzer Survey of Stellar Structure in Galaxies (S\textsuperscript{4}G) and its early-type galaxy extension are used to revisit the assessment of $c/a$ free from dust extinction and away from the influence of a stellar bulge.
We present a simple definition of $c/a$ and explore trends with other galactic physical parameters: total stellar mass, concentration index, total \HI mass, mass of the central mass concentration, circular velocity, model-dependent scales, as well as Hubble type.
Other than a dependence on early/late Hubble types, and a related trend with light concentration, no other parameters were found to correlate with the intrinsic flattening of spiral galaxies.
The latter is mostly constant with $\langle c/a \rangle = 0.124 \pm 0.001 \ (\text{stat}) \pm 0.033 \ (\text{intrinsic/systematic})$ and greater for earlier types.

\end{abstract}

\keywords{Galaxy disks, Galaxy dynamics, Galaxy kinematics, Galaxy properties, Galaxy structure}


\section{Introduction}
\label{sec:intro}

Stellar disks viewed edge-on display a distinct thickness characterized by an intrinsic flattening, $c/a$.
Knowledge and modeling of the latter is important for understanding galaxy structure \citep{Comeron2012,DiazGarcia2022}, accretion history \citep{TothOstriker1992, Buck2020, Hopkins2023}, galaxy evolution \citep{Kautsch2009,MengGnedin2021,GarciaMinchev2023}, constraining numerical simulations \citep{MengGnedin2021,Sotillo-Ramos2022, Hopkins2023}, and more. 
Beyond the physical understanding of galaxy flattening, the determination of $c/a$ directly affects galaxy inclinations, which are themselves critical for the deprojection of observables such as line widths, rotation curves, and light profiles, among others, from which global galaxy structural parameters may be estimated. 
For instance, estimates of galaxy mass and surface brightness profile decompositions, e.g., into the sum of a bulge and disk components, rely on suitable value(s) of $c/a$.

Historically, the determination of $c/a$ has employed a statistical analysis of the apparent $c/a$ distribution for large galaxy samples with all projected inclinations yielding an extrapolated intrinsic $c/a$ distribution \citep[e.g.,][]{Holmberg1946, Sandage1970, Guthrie1992, Lambas1992, Giovanelli1994, Unterborn2008}. 
This approach was typically justified by assuming that (i) disk galaxies are oblate spheroids and (ii) that the intrinsic $c/a$ distribution should be Gaussian. 
The former assumption is too restrictive \citep[disk galaxies are slightly triaxial;][]{Lambas1992, Fasano1993, Alam2002, Ryden2004} while the latter is fair.

Measurements of $c/a$ have also been ill-characterized, especially for older values based on photographic plates whose limiting isophotes were constrained by the sensitivity of the plates \citep{ReferenceCatalogue1964, UGC}.
Measurements of $a$ and $c$ have often been made at arbitrary locations \citep[see][for an in-depth discussion]{Trujillo2020}.
To remedy this issue, we define the location of $c$ relative to $a$ by way of ``characteristic cuts'', which consist of rigorous and reproducible measurements of $c/a$ for any disk galaxy observed with a digital detector (detailed in \sec{iso_cuts}).
Our study attempts to provide such a rigorous, straightforward, definition of $c/a$, free of extinction effects and the influence of a bulge, for accurate astrophysical applications as well as comparisons with numerical models of disk galaxies. 

Model-dependent axis ratios, like the ratio of disk scale height, $z$, to scale length, $h$, have occasionally been used in lieu of $c/a$ as a measure of true disk thickness \citep[e.g.,][]{Kregel2004,Padilla2008, Stark2009, vdWel2012, Mosenkov2015, Shibuya2015, Doore2021}.
This approach alleviates the need for a specific measurement location for $c$.
However, significant drawbacks of this approach are that the adopted model to estimate scale parameters, whether in length or height, may not be realistic and may be arbitrarily complex (e.g., given bars, breaks, lenses, and model parameterizations).
For instance, exponential disk models representing the radial light drop-off of luminous disks with parametric scale lengths, $h$,have often been invoked for disk galaxies \citep[e.g.,][]{Freeman1970,Kregel2002,Erwin2005,Hernandez2006,Erwin2008}.
However, offsets from this model can be significant \citep{Erwin2005, Pohlen2006, Erwin2008, Comeron2018, Mosenkov2022}.
Similar concerns hold for vertical profiles, whether exponential with scale height $z_0$, $\sech^2$ with scale height $z_{\sech^2} = 2z_0$ \citep[e.g.,][]{vdK1981,deGrijs1998,Yoachim2006,Mosenkov2015}, or S{\'e}rsic models \citep{vdWel2012}.
Given the many assumptions in any model-dependent fit, such practice requires great care in execution and application.

One-dimensional (1D) parametric decompositions targeted at retrieving intrinsic disk flattening require careful consideration of bulge light.
For example, \citet{deGrijs1998}'s 1D reduction reported $\langle z_0/h \rangle \sim 0.15$ for spiral galaxies, yet 1D photometric profiles were found to overestimate $z_0/h$ \citep{Kregel2002}, especially in galaxies with large bulges \citep{Mosenkov2015}.
Both \citet{Kregel2002} and \citet{Mosenkov2015} performed two-dimensional (2D) bulge-disk decompositions on spirals and reported $\langle z_0/h \rangle \sim 0.13$.
The three-dimensional (3D) SB profile decompositions of \citet{Mosenkov2022} found even thinner stellar disks with $\langle z_0/h \rangle \sim 0.1$ at $3.4 \ \mu$m.
In what follows, comparisons between our model-independent approach and model-dependent results will be presented. 

Our paper is organised as follows.
Our brief historical review in \sec{history} reveals that most previous measurements of $c/a$ have largely relied on ground-based blue sensitive, and thus dust-extinction prone, material and lacked a precise definition of the $c$ and $a$.
We suggest a method based on infrared digital surveys of disk galaxies. 
In \sec{data}, we present the data used in our analysis, ranging from science-ready images to disk decompositions from other works.
Our data reduction methodology to identify isophotes that characterise the disk's structure, as well as relevant statistical concerns, are addressed in \sec{autoprof}.
Results are presented in \sec{Flattening}. 
The shape of the intrinsic flattening distribution, and whether it is normal, is considered along with correlations between our (model-independent) $c/a$ values and structural galaxy properties, including similar measures from parametric (model-dependent) disk decompositions.
We also test for any biases due to galactic environments.
A discussion of our main findings is presented in \sec{Discussion}, including a discussion contrasting model-independent and model-dependent estimates of intrinsic flattening.
Future avenues are discussed in \sec{Future_Work} and followed by a summary in \sec{Conclusion}.
\app{AppendixA} discusses potential resolution effects on the measurement of $c/a$. 

\section{Historical Remarks} 
\label{sec:history}

Measurements of $c/a$ yield galaxy inclinations via Hubble's formula for oblate spheroids \citep{Hubble1926}:
\begin{equation}\label{eq:Hubble_form}
\cos^2 i = \frac{q^2 - (c/a)^2}{1 - (c/a)^2},
\end{equation}
where $i$ is the inclination, $q$ is the observed flattening, and
$a$ and $c$ are the semi-major and semi-minor axes of the disk, respectively.
In all but edge-on views, the intrinsic flattening is never observed.
The intrinsic thickness, $c/a$, of a galaxy disk is also ideally measured at a given surface density (brightness) level or fiducial location. 

Early estimates of $c/a$ relied on measuring apparent galaxy sizes off of photographic plates.
The latter were mostly sensitive to blue light and thus affected by dust extinction. 
\citet{Holmberg1946} developed a model to map the distribution of observed versus intrinsic axial ratios for idealized spiral galaxies whose true axis ratios and inclinations were known.
This model was then applied to a collection of 156 disk galaxies to infer an estimate for their intrinsic $c/a=0.2$ after assuming that $\cos i$ follows a uniform distribution.
The axial ratios $a$ and $c$ were measured visually at the observed galaxy edges by Holmberg and colleagues. 

\citet{Sandage1970} also studied the morphological dependence of $c/a$ for 168 elliptical, 267 lenticular, and 254 spiral galaxies imaged on {\it B}-band photographic plates.
These authors fitted various distribution functions to the observed distributions of randomly inclined sample galaxies in order to determine the intrinsic $c/a$. 
The semi-minor and semi-major axes for each galaxy were taken from the RC1 catalogue \citep{ReferenceCatalogue1964}. 
The Gaussian distribution of $c/a$ for both the spiral and lenticular galaxies had a mean $\langle c/a \rangle= 0.25$ and a deviation of $0.06$.

Other tabulations of $c/a$ for local disk galaxies include a private correspondence from B.M.~Lewis in \citet[][hereafter HG84]{Haynes1984}. 
Their Section IV (e) reports a distribution of $c/a$ ranging from $0.10$ for late spirals to $0.23$ for lenticulars. 
Details about the measurements of $c/a$ were not disclosed. 
Note that the tabulated values in \citetalias{Haynes1984} are reported and compared with our modern estimates in \Table{flattening_type_medians} below.

\citet{Guthrie1992} studied 888 isolated disk (Sa to Sc) galaxies from the Uppsala Catalogue of Galaxies \citep{UGC} to infer a distribution of intrinsic $\langle c/a \rangle = 0.11 \pm 0.03$.
As with earlier plate-based measurements, $a$ and $c$ were measured visually to the edge of each observed galaxy.
Guthrie asserted that these visual measurements correlated well with isophotal diameters at the 25 {\it B}-band \magss isophote.

Instead of considering a particular isophote, \citet{Lambas1992} defined the observed axial ratio using ellipticity values derived from density-weighted image moments, yielding a density-weighted apparent flattening.
These authors applied their method to a sample of 13,482 spiral, 4782 lenticular, and 2135 elliptical galaxies imaged in the {\it J}-band and digitized from plates by \citet{Loveday1989}.
Under the assumption that axis lengths follow a Gaussian distribution, they constructed a triaxial model for the spirals and lenticulars via Monte Carlo simulations.
Ultimately, \citet{Lambas1992} reported $\langle c/a \rangle \approx 0.2$ for spirals and $\langle c/a \rangle \approx 0.5$ for lenticulars.
Their large values were likely biased by stray light from a spheroidal component like a bulge.

\citet{Giovanelli1994} used digital {\it I}-band images of 1235 Sbc and Sc galaxies to measure $c/a$ at the 23.5 \magss isophote.
Their model of the observed flattening distribution, which does not depend on an explicit distribution function, produced a best fit $c/a$ value of $0.13$, however the authors suggest that the true mean $c/a$ ought to be $\sim$0.10.
No uncertainty on $c/a$ was quoted and their model's standard deviation is especially small.
It is interesting that \citet{Guthrie1992} and \citet{Giovanelli1994} reported similar values ($\langle c/a \rangle = 0.11-0.13$), despite the former relying on a magnifying glass, upon which a measuring scale was printed, and the latter using digital measurements to measure axis ratios.

More recent investigations on this topic are scanty, with a study by \citet{Unterborn2008} standing out.
They constructed a sample of 16,155 late-type spiral galaxies from the sixth data release of the Sloan Digital Sky Survey \citep{SDSS2000, SDSS2008}.
Measuring $c/a$ at the {\it r}-band 25 \magss isophote and following the statistical method of \citet{Sandage1970}, \citet{Unterborn2008} found $\langle c/a \rangle \approx 0.22$ for slightly triaxial disks.

Overall, published values of $c/a$ (intrinsic) for disk galaxies have ranged between 0.1 and 0.25.
We are revisiting measurements of $c/a$ since: 
(i) no previous study of $c/a$ relied on dust-free (transparent), sky-free, digital images of disk galaxies; 
(ii) a precise definition of $a$ and $c$ (or choice of a sensible distribution function) 
was seldom provided; 
(iii) formal errors on $c/a$ were rarely assessed. 

The definition of an unbiased estimate of the ``true'' flattening of disk systems is challenging since disks are not monolithic slabs; within the optical radius, the vertical height of a disk decreases (often exponentially) with galactocentric radius \citep[e.g.,][]{vdK1981, deGrijs1997, deGrijs1998, vdK2011, Bizyaev2014}, and may increase again in the outskirts due to disk flaring or warping \citep{Olling1995,OssaFuentes2023}.
Of issue then is the availability and widespread application of a rigorous definition of $a$ and $c$ which appropriately describes the disk.

Because parametric decompositions \citep{Kregel2004,Mosenkov2015,Erwin2015} are subjective and generally non-unique \citep{Knapen1991,MacArthur2003,Arora2021}, we favour direct measurements of $c/a$ for an assessment of intrinsic flattening from edge-on systems since surfaces of constant density, whether by light or stellar mass, can be measured and reported unambiguously.

With the availability of digital infrared surveys of the local galaxies, such as the volume-limited \emph{Spitzer} Survey of Stellar Structures in Galaxies (S\textsuperscript{4}G; \citealt{Sheth2010,MunosMateos2015,S4GDOI}), it becomes possible to study the intrinsic flattening, $c/a$, of disk galaxies and address the main issues above in a reliable and repeatable manner.  
The aim of the current study is to provide a straightforward non-parametric measurement of intrinsic flattening for any disk galaxy, based on reliable statistical methods and an extinction-free collection of edge-on galaxies.

\section{Data}
\label{sec:data}

\subsection{The Spitzer Survey of Stellar Structures in Galaxies Sample}\label{sec:S4G}

To assess the intrinsic flattening of galactic stellar disks, we have used images from the volume-limited S\textsuperscript{4}G and its extended sample of supplementary early-type galaxies ($T < 0$) \citep{Watkins2022}.
The latter extended sample enables a valuable representation of lenticular galaxies. 
The S\textsuperscript{4}G is a survey of 2352 galaxies within a radial distance of $\sim \! 40$ Mpc ($z \leq 0.01$) observed with the infrared array camera (IRAC; \citealt{Fazio2004}) onboard the \emph{Spitzer} Space Telescope \citep{Werner2004} at $3.6$ and $4.5 \ \mu \text{m}$ wavelengths. 
The S\textsuperscript{4}G is magnitude-limited with $m_{B\text{corr}} < 15.5$, and size-limited to angular scales greater than $1'$.
The S\textsuperscript{4}G's deep imaging reaches an AB surface brightness of $\mu_{3.6\mu \text{m}} \approx 26$ \magss.
This provides deep infrared views of the stellar disk (unmatched from the ground) and its mass distribution, largely unaffected by dust extinction \citep{Rix1993,Conroy2013,Courteau2014}.
The S\textsuperscript{4}G can be broadly viewed as an unbiased survey of local disk galaxies. 

For the sake of this study, the S\textsuperscript{4}G survey was further restricted to a sub-sample of visually selected systems, yielding a sub-sample of 141 edge-on S\textsuperscript{4}G disk galaxies. 
Details of that selection are found in \cite{Comeron2018}.
A further 8 galaxies were removed from this sub-sample, as described in \sec{iso_rejects}, resulting in a final sample size of 133 transparent edge-on S\textsuperscript{4}G galaxies for this study.

\begin{table}[]
    \centering
    \begin{tabular}{c|c|c|c}
        \toprule
       Hubble & Morphological & Hubble & \# of Sample \\
       Class & Classification & Type & Galaxies  \\
       \midrule
       \multirow{4}{*}{Lenticular Galaxies} & S0\textsuperscript{-} & -3 & 7 \\
        & S0\textsuperscript{$\circ$} & -2 & 23 \\
        & S0\textsuperscript{+} & -1 & 12 \\
        & S0/a & 0 & 8\\
        \midrule
       \multirow{9}{*}{Spiral Galaxies}& Sa & 1 & 1 \\
        & Sab & 2 & 5 \\
        & Sb & 3 & 4 \\
        & Sbc & 4 & 7 \\
        & Sc & 5 & 13 \\
        & Scd & 6 & 14 \\
        & Sd & 7 & 27 \\
        & Sdm & 8 & 7 \\
        & Sm & 9 & 5 \\
        \bottomrule
    \end{tabular}
    \caption{Morphological classifications of the S\textsuperscript{4}G galaxies used in this study, as per the CVRHS system, and number of galaxies of each morphology within our sample.}
    \label{tab:morphology_classifications}
\end{table}

S\textsuperscript{4}G galaxies were classified according to the CVRHS system by \citet{Buta2015}, with updated $T$-type classifications \citep[][]{Watkins2022}.
The morphological distribution of galaxies of each $T$-type (and corresponding morphology) in our sample is presented in \Table{morphology_classifications}.
It should be noted that galaxy classification of edge-on systems is especially problematic, though the reduced dust extinction in the IRAC bandpass is a significant improvement for this task over bluer dust-proned images.
Identifying barred systems in edge-on cases is also challenging, unless a peanut bulge is clearly seen. 
However, as seen in \sec{q_distribution}, barredness does not play a role in our assessment of $c/a$.

\subsection{Scale Parameters}
\label{sec:SBP}

As described below (\sec{iso_cuts}), our non-parametric method to estimate the intrinsic flattening, $c/a$, of galaxy disks relies on isophotal fitting of (transparent) galaxy images. 
Galaxy surface brightness (SB) profiles, obtained from radial cuts for edge-on systems, can also be used to measure relevant structural parameters such as the disk scale height, $z_0$, and scale length, $h$, and to isolate substructure. 
We will compare in \sec{Flattening} our non-parametric method of measuring $c/a$ to the parametric ratio of $z_0$ and $h$, the disk scale height and length.
To this end, we have used \citet{Comeron2018}'s decompositions of (major axis) SB profiles for our S\textsuperscript{4}G galaxies to source values of the scale heights for the thin and thick disks.  
We have also extracted disk scale lengths from the (major axis) SB profiles of our S\textsuperscript{4}G galaxies by fitting a single exponential to the disk-dominated regions of the profiles.

\subsection{Global Parameters}\label{sec:Pparam}

In forthcoming sections (\sec{q+other} and beyond), we will compare our estimates of $c/a$ with a selection of global physical galaxy parameters in order to explore any correlations.
These global parameters include the total stellar mass, $M_{*}$; the central light concentration index, $C_{31}$; the total \HI mass, $M_{\text{\HI\ }}$; the central mass concentration (CMC) mass, $M_{\text{CMC}}$; and the circular velocity, $V_{\text{rot}}$.
Because our galaxies are viewed edge-on, corrections to $V_{\text{rot}}$ for inclination were not applied. 

The stellar masses, $M_{*}$, were derived for the entire S\textsuperscript{4}G sample by \citet{MunosMateos2015}.
Their computation assumed a global stellar $M_*/L$ of $M/L_{3.6 \ \mu \text{m}} \approx 0.6 \ (M_\odot/L_\odot)$ \citep{Meidt2014}. 
The uncertainty in the adopted $M/L$ is large as estimates for $M/L$ ratios in this band range from $0.2-0.6 \ M_\odot/L_\odot$ (\citealt{Eskew2012}; \citealt{Conroy2013}; \citealt{Courteau2014}; \citealt{Cluver2014}; \citealt{Meidt2014}; \citealt{Hall2018}).
\citet{MunosMateos2015} also provided a concentration parameter $C_{31}$, which serves as a proxy for the bulge-to-disk luminosity ratio.
$C_{31}$ is a non-parametric structural parameter defined as the ratio between the radii that enclose 75\%, $r_{75}$, and 25\%, $r_{25}$, of the total light of a galaxy.

Our $M_{\text{\HI}}$ values were derived via the basic relation \citep{Giovanelli1998}
\begin{equation}\label{MHI}
\begin{aligned}
M_{\text{\HI}} &= 2.356 \times 10^5 D^2 \left[\frac{M_{\odot}}{\text{Mpc}^2 \ \text{Jy km s}^{-1}}\right] \int_v \frac{f_v \; \mathrm{d}v}{\left[\text{Jy km s}^{-1}\right]},\\
\phantom{M_{\text{\HI}}} &= 2.356 \times 10^5 D^2 10^{17.40 - m_{21,\text{c}}},
\end{aligned}
\end{equation}
where $D$ is the distance to the galaxy in Mpc, taken from \citet{MunosMateos2015}, and $m_{21,\text{c}}$ is the corrected magnitude of the 21 cm \HI line flux from \emph{HyperLeda} \citep{Makarov2014}.

\begin{figure}[ht!]
    \centering
    \includegraphics[width=\textwidth]{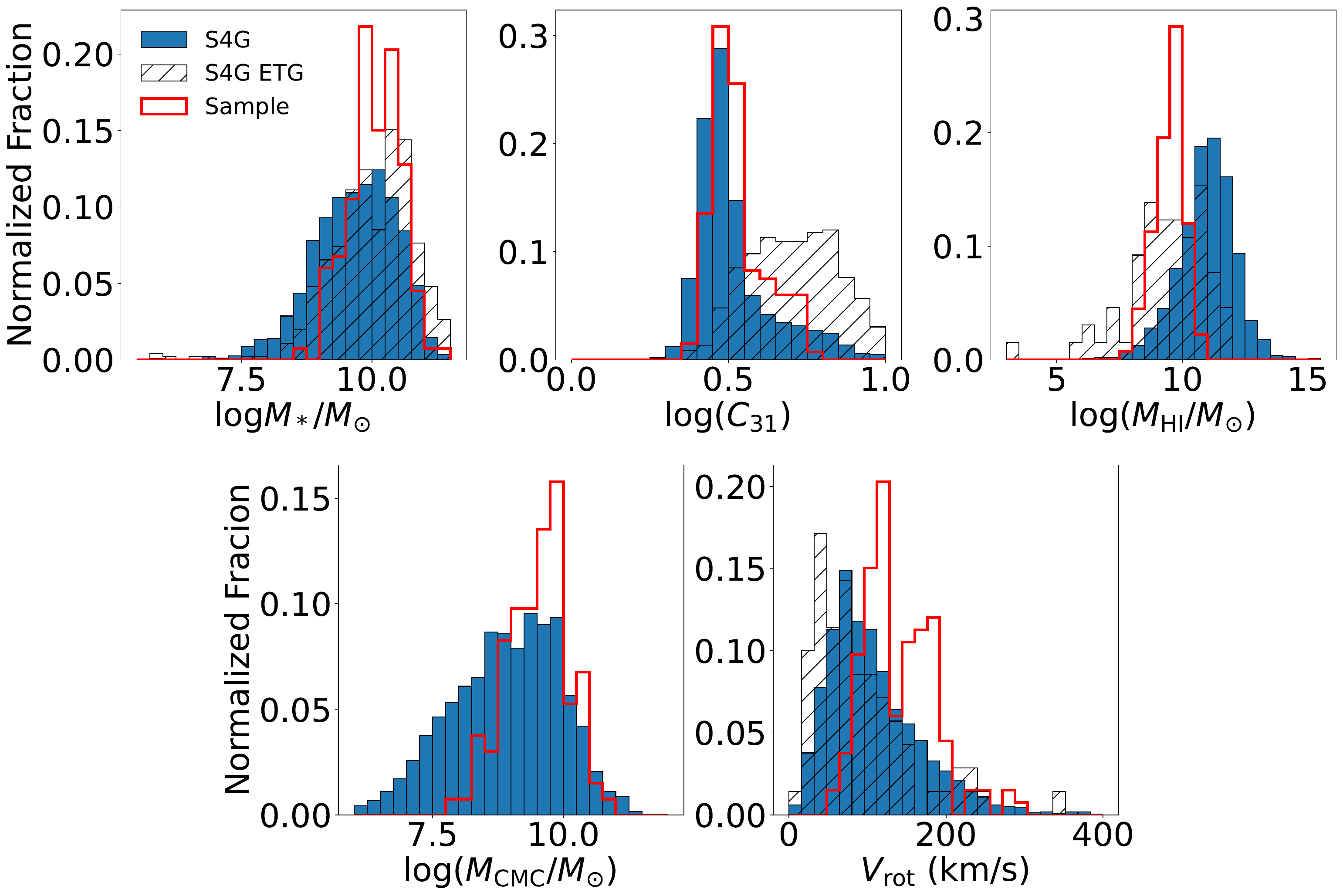}
    \caption{Comparison of our sub-sample (red histograms) with the S\textsuperscript{4}G catalogue (blue histograms) and (when available) the S\textsuperscript{4}G early type galaxy catalogue (hatched histograms).
    }
    \label{fig:sample_v_S4G_hist}
\end{figure}

\citet{Comeron2018} derived $M_{\text{CMC}}$, of which we avail ourselves, from their SB profile decomposition of S\textsuperscript{4}G galaxies.
Circular velocity ($V_{\text{rot}}$) data were also taken from the compilation of \citet{Comeron2018}.
If possible, \citet{Comeron2018} sourced data for the derivation of $V_{\text{rot}}$ from the Extragalactic Distance Database (EDD; \citealt{Tully2009}), which provides information and analysis of \HI\. line widths to determine $V_{\text{rot}}$.
Otherwise, \citet{Comeron2018} sourced $V_{\text{rot}}$ data from stellar absorption line studies (see \citealt{Comeron2018} for a comprehensive list of sources).
\Fig{sample_v_S4G_hist} shows a comparison of our sub-sample (red histograms) with the full S\textsuperscript{4}G sample.
Our sub-sample favors slightly more massive, though \HI\ poorer, systems.

Lastly, we consider whether the environments of our sample galaxies may affect their intrinsic flattening as a result of tidal interactions from neighbour galaxies, thus potentially leading to the heating of stellar disks. 
To investigate this effect, we sourced both the Dahari parameter \citep[$Q$;][]{Dahari1984} and the projected surface density to the third nearest neighbour galaxy from \citet{Laine2014}.

\section{Data Reduction Methodology: Surface Photometry}
\label{sec:autoprof}

Our definition of disk intrinsic flattening relies on the ratio of the characteristic radii $c$ and $a$.  
For an edge-on disk, $c$ represents the vertical axis above the disk, while $a$ is the semi-major radius measured at any surface brightness or density. 
While in principle incorrect for edge-on disks, we have applied 
elliptical isophote fitting to our galaxy images to extract $c/a$ at any galactocentric radius\footnote{A correct measurement of $c/a$ requires contours of constant density, which cannot be obtained without a full three-dimensional galaxy model as the vertical and axial galactic structures follow different functional forms.
Still, our application of selected cuts to edge-on systems provides an operational definition of $c/a$.}.

We have used the Python-based automated light profile extraction tool AutoProf \citep{Stone2021b} to perform image pre-processing and isophotal fitting on each S\textsuperscript{4}G sub-sample galaxy image.
An image pixel scale of 0.75\arcsec \ and a zero-point magnitude of 18.32 in the Vega system were used for our S\textsuperscript{4}G images.
Masks for each galaxy image were retrieved from the S\textsuperscript{4}G Pipeline 2 \citep{Sheth2010, MunosMateos2015} and used in AutoProf during processing.
Deconvolution of our images with the IRAC 3.6 $\mu m$ PSF was not necessary, as justified in \app{AppendixA}.
As a typical example, \Fig{IC0335_ellipse_fit} shows the isophotal fit of IC~0335.
Our rejection criteria, detailed in \sec{iso_rejects}, resulted in a final sample of 133 galaxies.

\begin{figure}
    \centering
    \includegraphics[width=\textwidth,height=\textheight,keepaspectratio]{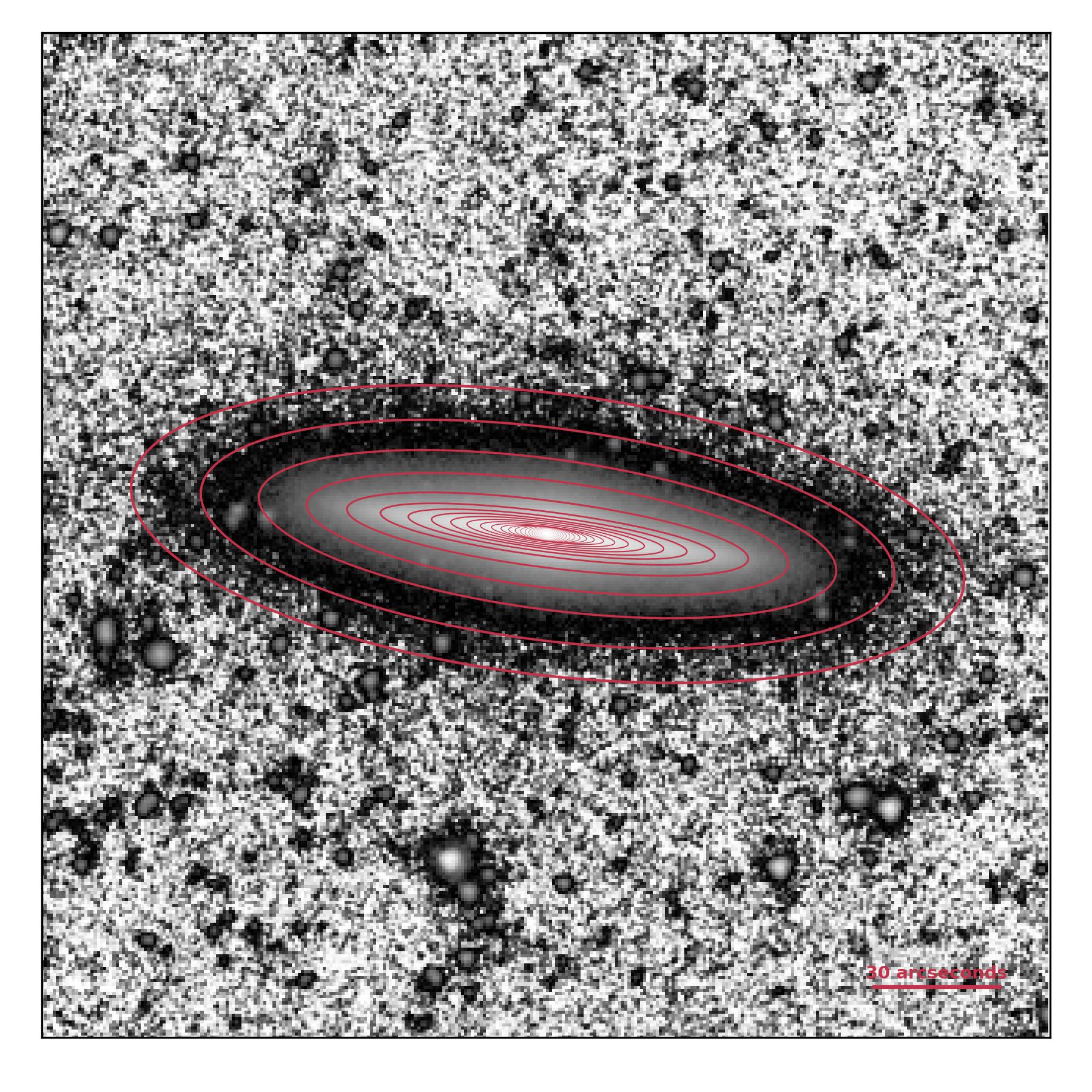}
    \caption{Contours of constant surface brightness (isophotes) for the galaxy 
    IC~0335 produced by AutoProf.
    Every fifth isophote is plotted as a red contour.}
    \label{fig:IC0335_ellipse_fit}
\end{figure}

\subsection{Characteristic Cuts}\label{sec:iso_cuts}

\begin{figure}
    \centering
    \includegraphics[width=0.85\textwidth,keepaspectratio]{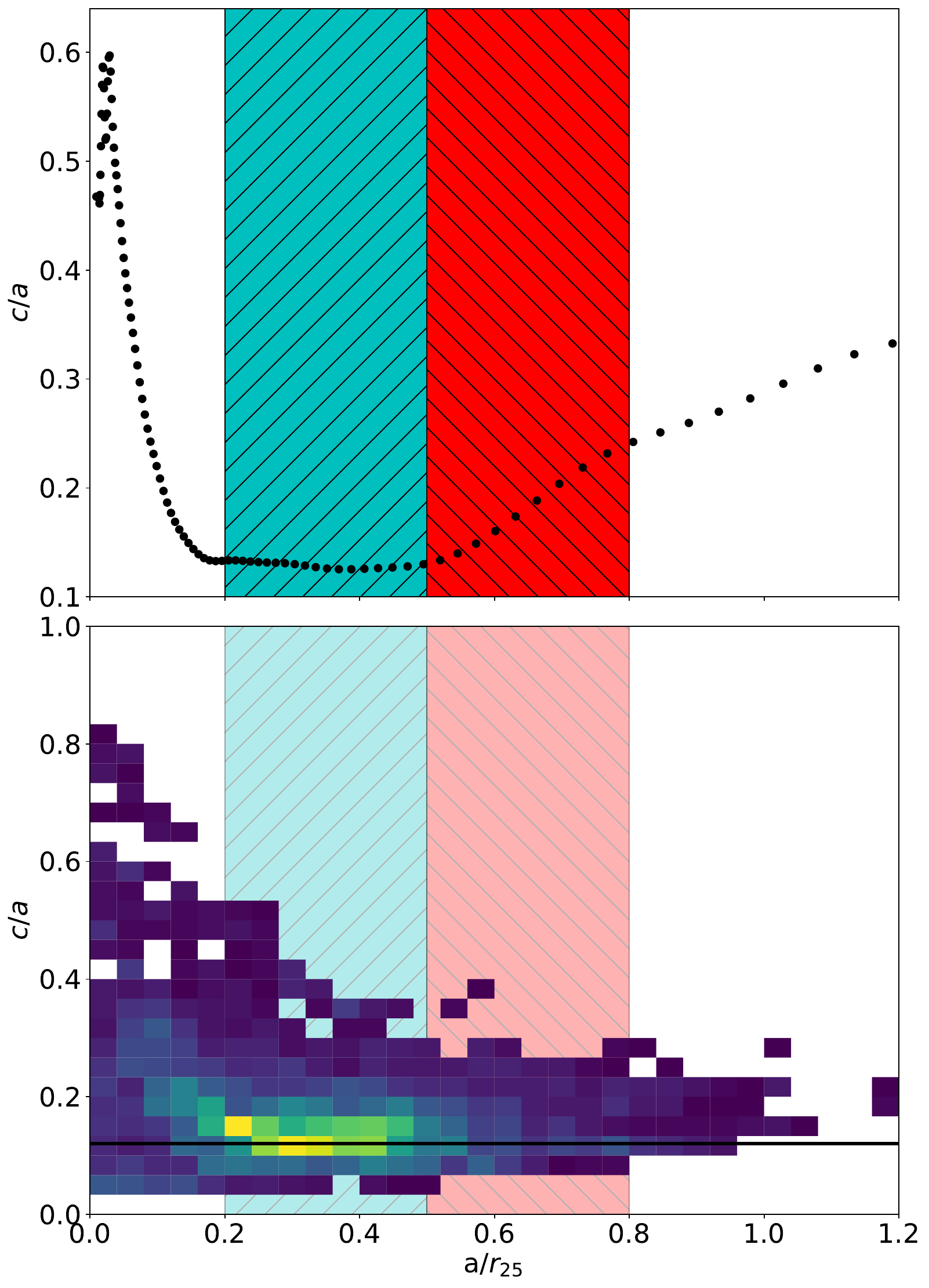}
    \caption{Top: Radial profile of intrinsic flattening for IC~0335 (as a typical example).
    The teal upward hatched area and the red downward hatched area define the inner cut from $0.2\!-\!0.5\,a/{r_{25}}$ and the outer cut over 
    $0.5\!-\!0.8\,a/{r_{25}}$, respectively. 
    Bottom: Stacked density distribution of all 133 intrinsic flattening radial profiles in our sample.
    The horizontal black line marks $c/a = 0.12$.
    See text for details.}
    \label{fig:IC0335_ellipticity}
\end{figure}

Our measurement of the intrinsic flattening of the stellar disk component used two ensembles of isophotes that span large sections of the disk. 
The constituent isophotes are those that intersect the galactic midplane within two cuts defined relative to the radius of the $25^{\rm{th}}$ B-band magnitude isophote, ${r_{25}}$.
These cuts, depicted in \Fig{IC0335_ellipticity}, are defined by $0.2\,{r_{25}} \leq |a| \leq 0.5\,{r_{25}}$, for the inner cut, and $0.5\,{r_{25}} \leq |a| \leq 0.8\,{r_{25}}$, for the outer cut where $a$ is the galactocentric radius along the major axis of the galaxy.
While this choice is somewhat arbitrary, they match those of \citet{Comeron2018}, allowing for direct comparisons.
These cuts also roughly divide isophotes dominated by the thin and thick disk light, respectively.

At radii below $0.2\,{r_{25}}$, rapid variations of $c/a$ characteristic of the bulge are found.
These should not enter our estimates of the intrinsic thickness of the disk. 
Within the inner cuts between $0.2 \! - \! 0.5\,{r_{25}}$, a sequence of lowest $c/a$ for the thin disk is expected, while higher values are expected for the outer cuts for the thick disk between $0.5 \! - \! 0.8\,{r_{25}}$. 
At larger radii beyond $0.8\,{r_{25}}$, $c/a$ should converge to any value of $c/a$ representative of the final galaxy outskirts, possibly affected by flaring or interactions with neighbours. 
Ideally, our measurements of $c/a$ ought to be made within the inner cuts as we demonstrate below. 
Furthermore, concerns about light scattered by PSF wings are alleviated in the inner cuts as these are regions of relatively high surface brightness.
For the dimmer outer cuts, scattered light from the midplane was also shown to be small \citep{Comeron2018}.

The mean $c/a$ of isophotes within both inner and outer cuts is taken to be the intrinsic flattening within the region of the respective cut. 
The error on this mean is the quadrature sum of the standard deviation of $c/a$ values within the cut and the uncertainty resulting from the propagation of each individual isophote’s $c/a$ error through the calculation of the mean.
The ellipticity of the isophotes within a given cut varies for all galaxies, so this averaging smooths over any fluctuations in thickness, allowing us to probe the true flattening of the disk. 
As an example, the top panel of \Fig{IC0335_ellipticity} shows the radial variations of $c/a$ for the galaxy IC~0335.
While no two galaxies have identical radial $c/a$ profiles, IC~0335 highlights our method's ability to sample two distinct disk structural regimes.
The bottom panel of \Fig{IC0335_ellipticity} presents the density distribution of the $c/a$ profiles with radius for all 133 galaxies in our sample.
Our choice of cuts is meant to separate the CMC and the thin/thick disk components.

We can divide our sample into two global sub-samples: the first consisting of all spiral galaxies, and the second of all lenticular galaxies in our sample. 
The black line at $c/a = 0.12$ in the bottom panel of \Fig{IC0335_ellipticity} anticipates results from our analysis of the distribution of $c/a$ for our spiral sub-sample.
This line describes the mean inner cut $c/a$ of spiral galaxies and highlights that the outer cuts are systematically thicker than the inner cuts, as expected.

\subsection{Rejected Isophotal Fits}
\label{sec:iso_rejects}

Despite AutoProf’s robust data reduction, we visually reviewed the resulting isophotal fits to ensure no galaxies with erroneous isophotes entered our analysis. 
Isophotes in the inner and outer cuts were expected to be well behaved, as in the top panel of \Fig{IC0335_ellipticity}.
All together, three fits were affected by foreground objects and rejected.
AutoProf failed to fit a further five galaxies with masking, yielding a final sample of 133 galaxies.

\subsection{Statistical Corrections}\label{sec:math_methods}

As a result of our inability to choose the exact viewing angle of objects in the sky, we consider corrections for deviant galaxy orientations within our data set.
\sec{inclination} explores reasonable and worst-case scenarios for a galaxy viewed almost edge-on, while \sec{tri-ax} derives the statistical correction required to account for the random azimuthal rotation of galaxies.

\subsubsection{Inclination Effects}\label{sec:inclination}

In order to observe the true intrinsic flattening of a disk, it must be viewed exactly edge-on at $i = 90^\circ$.
However, the reported inclinations (from \emph{HyperLeda}) of our sample fall within $78^\circ \lesssim i \leq 90^\circ$.
These inclinations were derived using \Eq{Hubble_form} and are therefore biased by previous measurements of intrinsic flattening.
Visual inspection (by the main author as well as \citealt{Comeron2012}, \citealt{Mosenkov2015}, and \citealt{Comeron2018}) suggests that the galaxies in our sample are viewed almost perfectly edge-on.
We disregarded previously reported inclinations to eliminate any bias. 
For robust independent measurements of $c/a$, we accepted that some disks may be ``excessively" thick and trust our visual pattern recognition \citep[e.g., \emph{Galaxy Zoo},][]{Lintott2008}.
We posit that the true inclinations of the galaxies in our sample fall within the range $85^\circ < i < 90^\circ$.

In light of this, one may ask to what degree viewing a galactic disk at inclinations less than $90^\circ$ biases our measurement of $c/a$?
Note that a non-zero tilt will always increase the observed thickness, $q$ in \Eq{Hubble_form}, or equivalently $c_{\rm{obs}}/a_{\rm{obs}}$.
In this case, the maximum error will occur at $i = 85^\circ$.
If we view a transparent disk with $c/a = 0.13$ at this inclination, then by solving for $q$ in \Eq{Hubble_form} we expect to overestimate $c/a$ by $\sim \! 20\%$.
Likewise, a disk with $c/a = 0.20$ at this inclination will result in an overestimate of $\sim \! 8\%$.
So, for measured $c/a$ values in the expected range, we will, at worst, underestimate the flattening by $\lesssim 0.02$.

Since the sample is heavily biased against inclinations that deviate from the true edge-on perspective \citep{Comeron2018}, we expect an overestimated $c/a$ by $\sim \! 1\%$.
In all but the largest measured $c/a$ values, these overestimates are on the order of the systematic noise.
So, $c/a$ outliers resulting from deviant inclinations are expected to be rare, and measurements are not corrected for this effect.
Nevertheless, to minimize the potential effect of a small number of galaxies that deceptively appear to be aligned edge-on, we also used the median statistic of $c/a$ to minimize the impact of any outliers in our analysis.

\subsubsection{Triaxial Variance Effects}\label{sec:tri-ax}

Galactic disks in spiral galaxies are triaxial spheroids obeying $c < b < a$ (rather than $c < b = a$).  
On average, spiral galaxies are only slightly triaxial, with mean face-on axial ratio $\langle b/a \rangle = 0.9$ \citep{Lambas1992}. 
Due to the random orientation of galaxies in the sky, measurements of $a_{\rm{obs}}$ have $b \leq a_{\rm{obs}} \leq a$.
On average, we expect $\langle a_{\rm{obs}} \rangle \simeq 0.97a$.
Therefore, we apply a 3\% downward correction to reported mean and median $c/a$ values. 
Uncorrected values are denoted by prime indices (e.g., $\mu'$).

\begin{figure}
    \centering
    \includegraphics[width=0.8\linewidth]{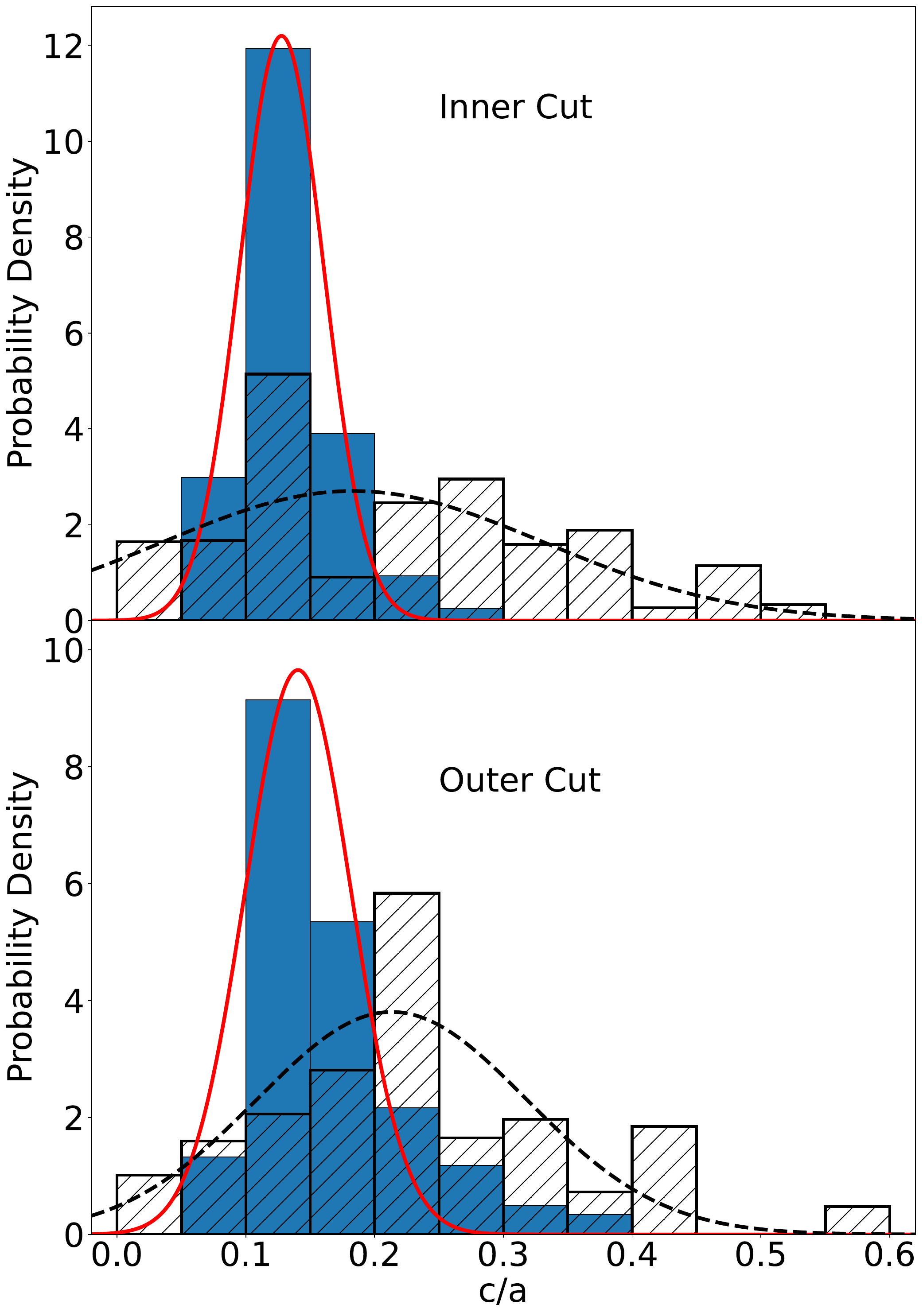}
    \caption{Intrinsic flattening distribution of the inner cut (top) and outer cut (bottom).
    The solid blue and hatched histograms represent spiral and lenticular galaxies, respectively. 
    The solid red and dashed black lines are Gaussian fits to the spiral distributions and the lenticular distributions, respectively. 
    }
    \label{fig:c/a_hist_fit}
\end{figure}

\section{Characterising Intrinsic Flattening in Edge-On Systems}
\label{sec:Flattening}

With our measurements of line-of-sight fluxes at well-defined galactocentric locations for $133$ S\textsuperscript{4}G galaxies, we can establish the distributions of intrinsic flattening, $c/a$, for spiral and lenticular galaxies (\sec{q_distribution}).
Their uncertainty-weighted histograms are presented in \Fig{c/a_hist_fit}.
We then explore the connection between disk thickness and morphology in \sec{q_distribution}, and test in \sec{q+other} for correlations between the physical parameters detailed in \sec{Pparam} and intrinsic flattening.
Our $c/a$ values are also compared to the parametric decompositions of \citet{Comeron2018} in \sec{ComeronCompare}.
We close this section with an assessment of environmental effects on disk thickness in \sec{environments}.

\subsection{The Distribution of Intrinsic Flattening vs Morphology}
\label{sec:q_distribution}

\begin{table}
    \centering
    \begin{tabular}{c|c|c|c|c}
        \toprule
          & \multicolumn{2}{c}{Spiral Galaxies} & \multicolumn{2}{c}{Lenticular Galaxies}  \\
         \cmidrule(lr){2-3}\cmidrule(lr){4-5}
          & Inner Cut & Outer Cut & Inner Cut & Outer Cut \\
         $\mu'$ & $0.128 \pm 0.001$ & $0.140 \pm 0.003$ & \phantom{-} $0.18 \pm 0.03$ & \phantom{-} $0.21 \pm 0.02$ \\
         $\mu$ & $0.124 \pm 0.001$ & $0.136 \pm 0.003$ & \phantom{-} $0.17 \pm 0.03$ & \phantom{-} $0.20 \pm 0.02$ \\
         $\sigma$ & $0.033 \pm 0.001$ & $0.041 \pm 0.003$ & \phantom{-} $0.15 \pm 0.02$ & $\phantom{-} 0.10 \pm 0.02$ \\
         \midrule
         $m$ & $0$ & $0$ & $-0.04 \pm 0.02$ & $-0.01 \pm 0.01$ \\
         $b$ & $0.141 \pm 0.005$ & $0.16 \pm 0.01$ & $\phantom{-}0.18 \pm 0.03$ & $\phantom{-}0.20 \pm 0.02$ \\
         \bottomrule
    \end{tabular}
    \caption{Top section: Mean and standard deviation of the Gaussian fits performed on the intrinsic flattening distributions of \Fig{c/a_hist_fit}.
    $\mu'$ is the mean without accounting for triaxiality (see \sec{tri-ax}), while $\mu$ includes this correction.
    Bottom section: Slope, $m$, and y-intercept, $b$, of the linear fits mapping $T$ to $c/a$ (plotted in \Fig{flattening_by_T}) with the functional form $c/a = mT + b$. 
    The slopes for spiral galaxies are statistically consistent with zero.
    }
    \label{tab:gaussian_params}
\end{table}

\begin{table}
    \centering
    \begin{tabular}{c|c|c|c|c}
        \toprule
        Hubble & $N$\textsuperscript{\emph{a}} & Inner Cut & Outer Cut & HG84 \\
        Type & \ & Median c/a & Median c/a & $c/a$ \\
        \midrule
        $-3$ & $7$ & $0.27 \pm 0.15$ & $0.20 \pm 0.16$ & 0.23 \\
        $-2$ & $23$ & $0.28 \pm 0.12$ & $0.22 \pm 0.10$ & 0.23 \\
        $-1$ & $12$ & $0.23 \pm 0.11$ & $0.21 \pm 0.09$ & 0.23 \\
        $\phantom{-}0$ & $8$ & $0.15 \pm 0.17$ & $0.18 \pm 0.13$ & 0.23 \\
        $\phantom{-}1$ & $1$ & $0.146 \pm 0.003^b$ & $0.131 \pm 0.001^b$ & 0.23 \\
        $\phantom{-}2$ & $5$ & $0.12 \pm 0.07$ & $0.12 \pm 0.10$ & 0.20 \\
        $\phantom{-}3$ & $4$ & $0.18 \pm 0.05$ & $0.20 \pm 0.06$ & 0.175 \\
        $\phantom{-}4$ &  $7$ & $0.14 \pm 0.04$ & $0.18 \pm 0.05$ & 0.14 \\
        $\phantom{-}5$ & $13$ & $0.14 \pm 0.04$ & $0.13 \pm 0.04$ & 0.103 \\
        $\phantom{-}6$ & $14$ & $0.13 \pm 0.02$ & $0.14 \pm 0.04$ & 0.10 \\
        $\phantom{-}7$ & $27$ & $0.14 \pm 0.04$ & $0.17 \pm 0.07$ & 0.10 \\
        $\phantom{-}8$ & $7$ & $0.12 \pm 0.04$ & $0.14 \pm 0.05$ & 0.10 \\
        $\phantom{-}9$ & $5$ & $0.15 \pm 0.04$ & $0.19 \pm 0.05$ & 0.10 \\
        \bottomrule
    \end{tabular}
    \caption{Median intrinsic flattening values for the inner and outer cuts versus Hubble Type $T$.
    The reported errors are the standard deviations of the mean.
    \emph{a} - Number of galaxies.
    \emph{b} - The uncertainty is a result of a singular data point and is grossly underestimated.
    The last column shows the $c/a$ values reported in \citetalias{Haynes1984}.}
    \label{tab:flattening_type_medians}
\end{table}

\begin{figure}
    \centering
    \includegraphics[width=\linewidth]{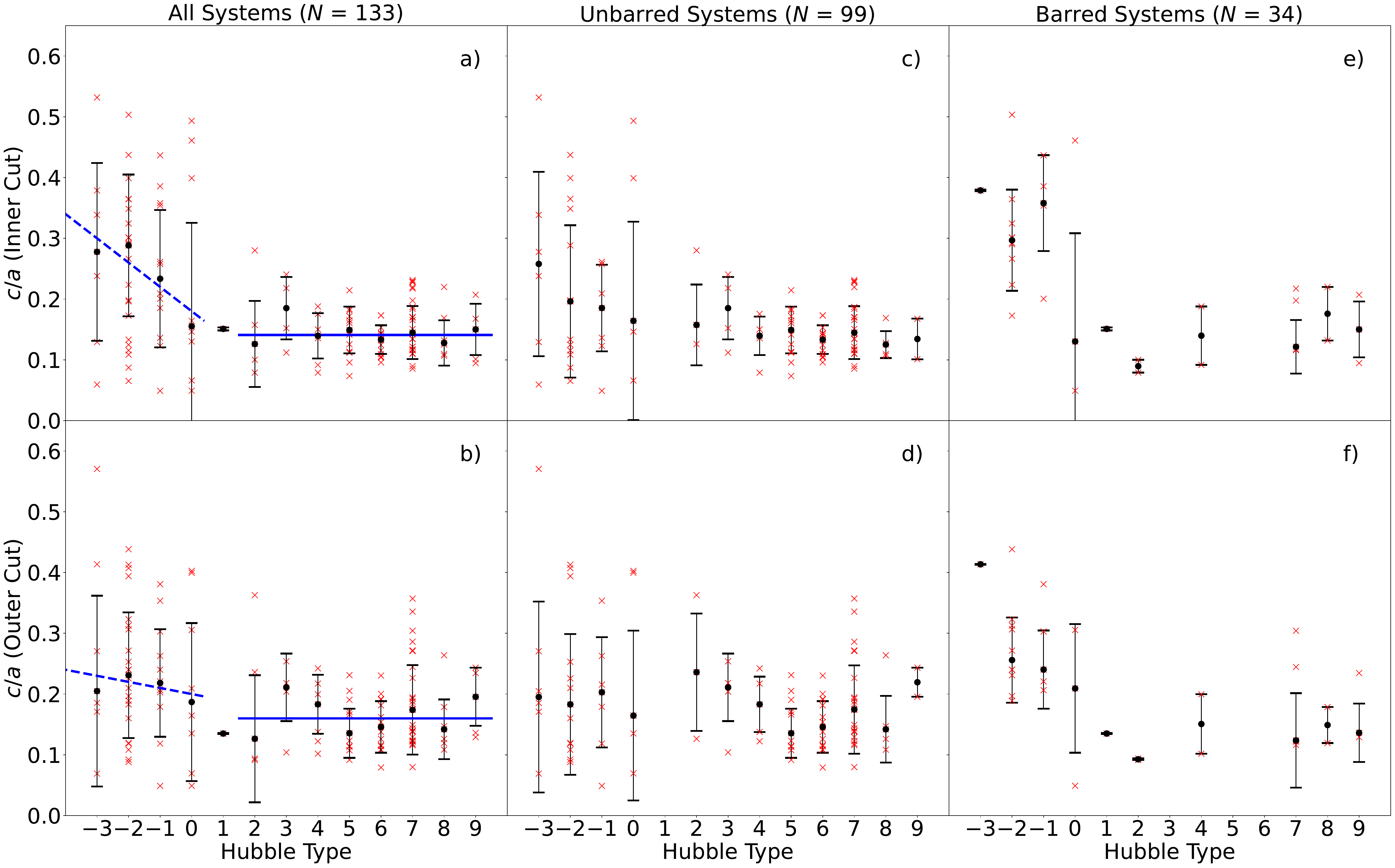}
    \caption{Intrinsic flattening of disk galaxies by Hubble Type based on the inner (top) and outer (bottom) cuts.
    Black points are the median $c/a$ values for each Hubble Type, while error bars represent the dispersion. 
    Red \emph{x}s are $c/a$ values for individual galaxies. 
    Error bars on the individual galaxy points have been omitted for the sake of clarity and are on the order of 1-2\%.
    In panels a)-b), each solid blue line is a linear fit to the median flattening of disk galaxies ranging from $T$ = 2 to $T$ = 9, while each dashed blue line is a linear fit to the lenticular galaxies median $c/a$.
    In panels c)-f), unbarred and barred systems are shown separately without fits for visual comparison with the full set.
    The $c/a$ distributions for barred and unbarred disk galaxies are comparable.}
    \label{fig:flattening_by_T}
\end{figure}

Our binning of galaxies into lenticular ($-3 \leq T \leq 0$) and spiral ($1 \leq T \leq 9$) morphologies reveals that spirals are well described by a Gaussian distribution for both the inner and outer cuts; see \Fig{c/a_hist_fit}.
The outer cut distribution of lenticulars in \Fig{c/a_hist_fit} is also reasonably Gaussian.
The mean and standard deviation of each Gaussian fit are presented in \Table{gaussian_params}. 

The spiral galaxies follow relatively tight $c/a$ distributions in both cuts, 
the narrowest width being achieved for the inner cuts. 
For the inner cuts of spiral galaxies in our sample, we find $\langle c/a \rangle = 0.124 \pm 0.033$.
This value for the intrinsic flattening agrees well with \citet{Guthrie1992} and \citet{Giovanelli1994}.

A small tail extending to a flattening of 0.4 exists in the outer cut spiral distribution. 
Since this tail is not present in the spiral’s inner cut distribution, it is unlikely to be an artifact of either inclination or triaxiality effects.
Rather, as discussed in \sec{iso_cuts}, this tail likely arises from slight luminous contributions from non thin-disk sources such as a thick disk, disk flaring, nearby objects, or even light scattered by the PSF \citep{Sandin2014, Sandin2015}.

For lenticular galaxies, the $c/a$ means for the inner and outer cuts (see \Table{gaussian_params}, as well as both panels of \Fig{c/a_hist_fit}) agree, while the deviations do not.
Both values are notably larger than their spiral counterparts.
This is a clear reflection of their more troubled past relative to much thinner spiral galaxies (see below). 

We report the median intrinsic flattening for each Hubble Type in \Table{flattening_type_medians}, along with the $c/a$ values reprinted in \citetalias{Haynes1984} for the corresponding $T$.
The individual data used to assemble \Table{flattening_type_medians} are also plotted in \Fig{flattening_by_T}, where the $c/a$ value of each galaxy is shown versus the galaxy's Hubble Type.
Measurements for each galaxy are marked as red \emph{x}'s in this figure.
The median $c/a$ values reported in \Table{flattening_type_medians} are also plotted as black points with error bars that denote the scatter of $c/a$ for a given $T$.
Panels a), c), and e) show the inner cut $c/a$ values, while b), d), and f) use outer cut values.

For lenticular galaxies ($T$ = [$-3$, 0]), we find that earlier systems appear to be thicker, albeit with large scatter, in agreement with older findings \citep{Binney1981}.
The thickening towards earlier types can be attributed to the evolutionary processes that shaped them; galaxy interactions increase the velocity dispersion in the disk while gas depletion halts the formation of new, low velocity dispersion stars \citep{Comeron2016, Pinna2019a, Pinna2019b}.
Our result above contrasts with \citetalias{Haynes1984}'s constant intrinsic flattening ($c/a = 0.23$) for all lenticulars. 

For the later-type disks ($T$ = [2, 9]), the median thickness is mostly constant with $c/a \simeq 0.14$.
This result holds within uncertainties for either inner or outer cuts.
\citetalias{Haynes1984} also quoted a nearly constant flattening for $T$ = [5, 9] disk galaxies, albeit with a smaller amplitude ($c/a \simeq 0.10$).
We find no evidence of the disk thickening reported by \citetalias{Haynes1984} from $T = 5$ to $T = 1$, though our counts for these types are low.

Panels a) and b) of \Fig{flattening_by_T} show linear fits to the lenticular and $T >$ 1 spiral $c/a$ values.
These fits are parameterised in \Table{gaussian_params}.
$T$ = 1 has been excluded from the spiral fits given the bin consists of one object.
The slopes of the spiral galaxy fits are statistically consistent with, and here fixed to, zero.

We further divided our sample into galaxies visually identified as tentatively hosting a bar or not.
The inner and outer cut $c/a$ values for unbarred galaxies are plotted against $T$ in panels c) and d) of \Fig{flattening_by_T}, respectively.
Given the prevalence of barred systems in local spiral galaxies \citep[e.g.,][]{Buta2015, Vera2016, Vazquez-Mata2022}, many bars have likely been missed due to the restricted edge-on views.
Panels c) and d) of \Fig{flattening_by_T} may therefore include a large number of unidentified barred galaxies.
Still, based on the available information and given the caveats noted above, no marked difference in the intrinsic flattening of barred vs unbarred galaxies is noticed. 

\subsection{Intrinsic Flattening and Other Structural Parameters} 
\label{sec:q+other}

\begin{figure}
    \centering
    \includegraphics[width=\linewidth]{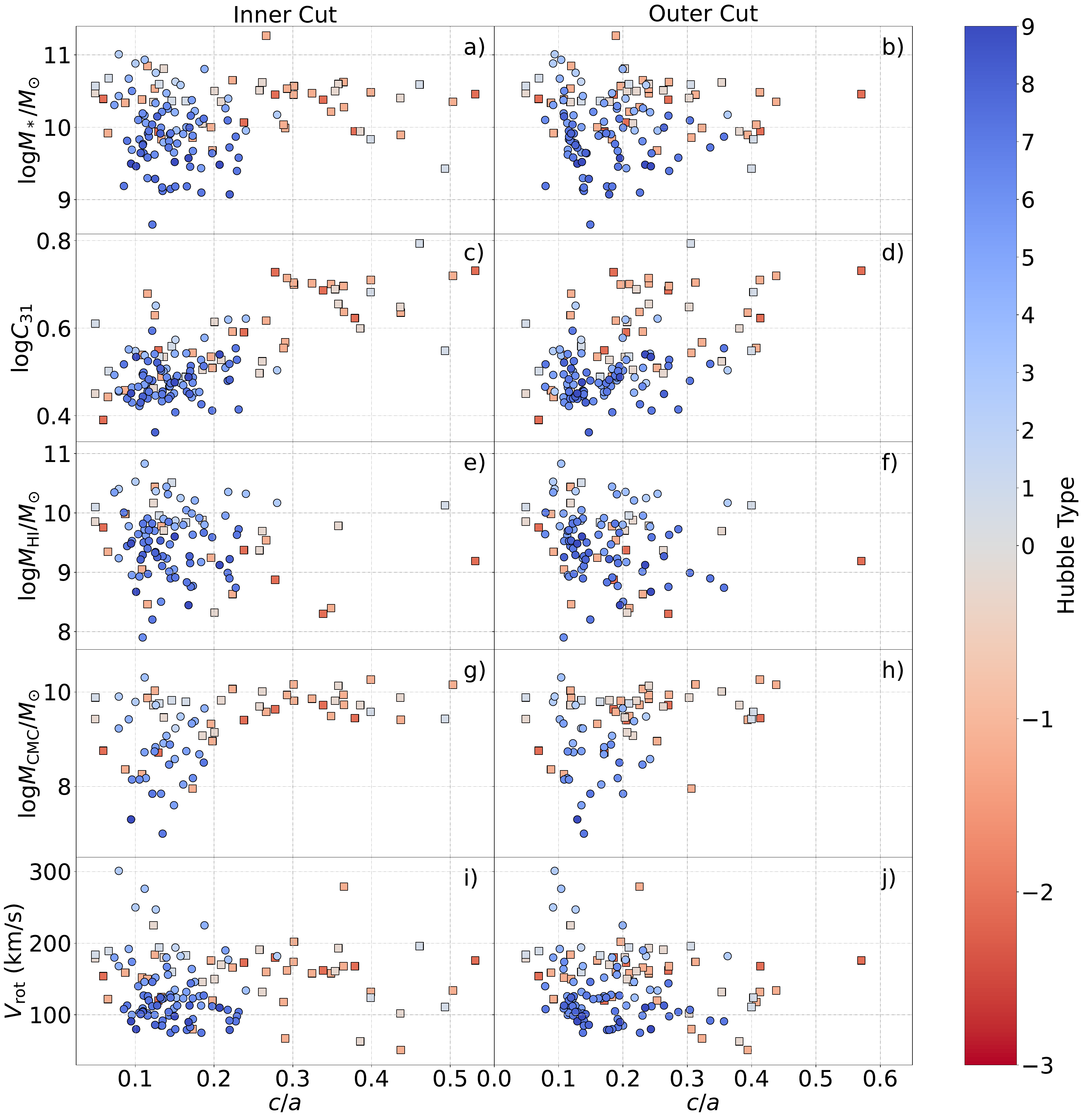}
    \caption{Various physical parameters plotted against intrinsic flattening $c/a$.
    From top to bottom, the physical parameters are; total stellar mass; concentration parameter, $C_{31}$; total \HI mass; CMC mass; and circular velocity.
    The left and right columns use the intrinsic flattening of the inner and outer cuts, respectively.
    Squares and circles represent lenticular and spiral galaxies, respectively.
    Colours scale with the Hubble Type of the galaxy.}
    \label{fig:other_params}
\end{figure}

We now investigate potential correlations between intrinsic flattening and other physical parameters listed in \sec{Pparam}.
Note that not all galaxies have recorded values for each parameter.
\Fig{other_params} shows those correlations with squares and circles denoting lenticulars and spiral galaxies, respectively.
The shade of coloured labels identifies the Hubble Type.
$T$ = 0 galaxies are grey squares.
The Pearson correlation coefficients for each relation shown in \Fig{other_params} are presented in \Table{other_params_pearson_r}.
A few significant correlations with $c/a$ can be gleaned from \Fig{other_params}, while most of the tested variables such as stellar mass, \HI mass, and circular velocity, show no correlation. 
In the next section, we examine the surprising lack of correlation between $c/a$ and $V_{\rm{rot}}$ in the context of parametric decompositions. 

\begin{table}
    \centering
    \begin{tabular}{c|c|cccc}
        \toprule
          & & \multicolumn{2}{c}{Inner Cut} & \multicolumn{2}{c}{Outer Cut} \\
          \cmidrule(lr){3-4}\cmidrule(lr){5-6}
          & $N$ & $r_p$ & $p$ & $r_p$ & $p$ \\
          \cmidrule(lr){2-2}\cmidrule(lr){3-3}\cmidrule(lr){4-4}\cmidrule(lr){5-5}\cmidrule(lr){6-6}
         $\log{M_{*}/M_\odot}$ & 133 & $\phantom{-}0.14$ & $\mathbf{10\%}$ & $\phantom{-}0.01$ & $\mathbf{87\%}$ \\
         $\log{C_{31}}$ & 133 & $\phantom{-}0.71$ & $\ll 1\%$ & $\phantom{-}0.49$ & $\ll 1\%$ \\
         $\log{M_{\text{\HI}}/M_\odot}$ & 106 & $-0.13$ & $\mathbf{19\%}$ & $-0.17$ & $\mathbf{8\%}$ \\
         $\log{M_{\rm{CMC}}/M_\odot}$ & 75 & $\phantom{-}0.45$ & $\ll 1\%$ & $\phantom{-}0.31$ & $0.6\%$ \\
         $V_{\rm{rot}}$ & 133 & $\phantom{-}0.03$ & $\mathbf{77\%}$ & $-0.14$ & $\mathbf{9\%}$ \\
        \bottomrule
        
    \end{tabular}
    \caption{Pearson correlation coefficients and their significance for the correlations between $c/a$ and the physical parameters plotted in \Fig{other_params}.
    $N$ is the number of galaxies for which the relevant parameter exists.
    $r_p$ is the Pearson correlation coefficient.
    $p$ is the two-sided $p$-value that denotes the significance of the correlation.}
    \label{tab:other_params_pearson_r}
\end{table}

The light concentration index, $C_{31}$, seems to correlate fairly well with $c/a$ for the inner cut ($r_p=0.71$), as quantified in \Table{other_params_pearson_r}.
The spiral and lenticular galaxies seem to form a global $c/a - \log{C_{31}}$ trend.
Closer investigation of this trend reveals that it is entirely supported by the lenticulars, while the spirals show no correlation\footnote{Caution is advised in interpreting any trend between $c/a$ and $\log{C_{31}}$ if the missing low values of $C_{31}$ at high $c/a$ for the lenticulars are due to selection effects.}.
Such a trend should also appear between $\log{M_{\rm{CMC}}}$ and $c/a$, but is surprisingly absent.
The $\log{M_{\rm{CMC}}}-c/a$ relations in panels g) and h) of \Fig{other_params} reveal boundaries defined by the least massive CMCs at any given $c/a$.
The spirals have a sharply rising lower bound on $M_{\rm{CMC}}$, while the lenticulars have central mass concentrations more massive than $10^9 \ M_{\odot}$, regardless of $c/a$.

Despite the lack of correlation of $c/a$ with many variables, similar boundaries stand out in the other panels of \Fig{other_params}.
For instance, panels a) and b) suggest a boundary for the minimum total stellar mass as a function of $c/a$.
For spirals, that boundary grows as the disk thickens (growing $c/a$), while lenticulars are limited to $\log (M_*/M_\odot) \geq 9.8$.
Similar boundaries are present in the other panels for spiral galaxies.

The boundaries in panels a), b), g) and h) of \Fig{other_params} for lenticular galaxies are largely driven by selection biases stemming from our requirement for highly inclined galaxies; bulges must be massive for their host to be identified as a lenticular in the edge-on perspective.
The S\textsuperscript{4}G includes lenticulars with stellar masses down to about $10^8 \ M_{\odot}$ \citep{Watkins2022}, confirming that our sample is deficient in lenticulars with stellar masses below $\sim \! \! 10^{9.8} \ M_{\odot}$.
It is therefore likely that the trend seen in the $\log{C_{31}}-c/a$ relation established by lenticular galaxies is mostly a result of selection. 

\subsection{Correlations between \textit{c/a} and Model-Dependent Parameters}
\label{sec:ComeronCompare}

\begin{figure}
    \centering
    \includegraphics[width=\linewidth]{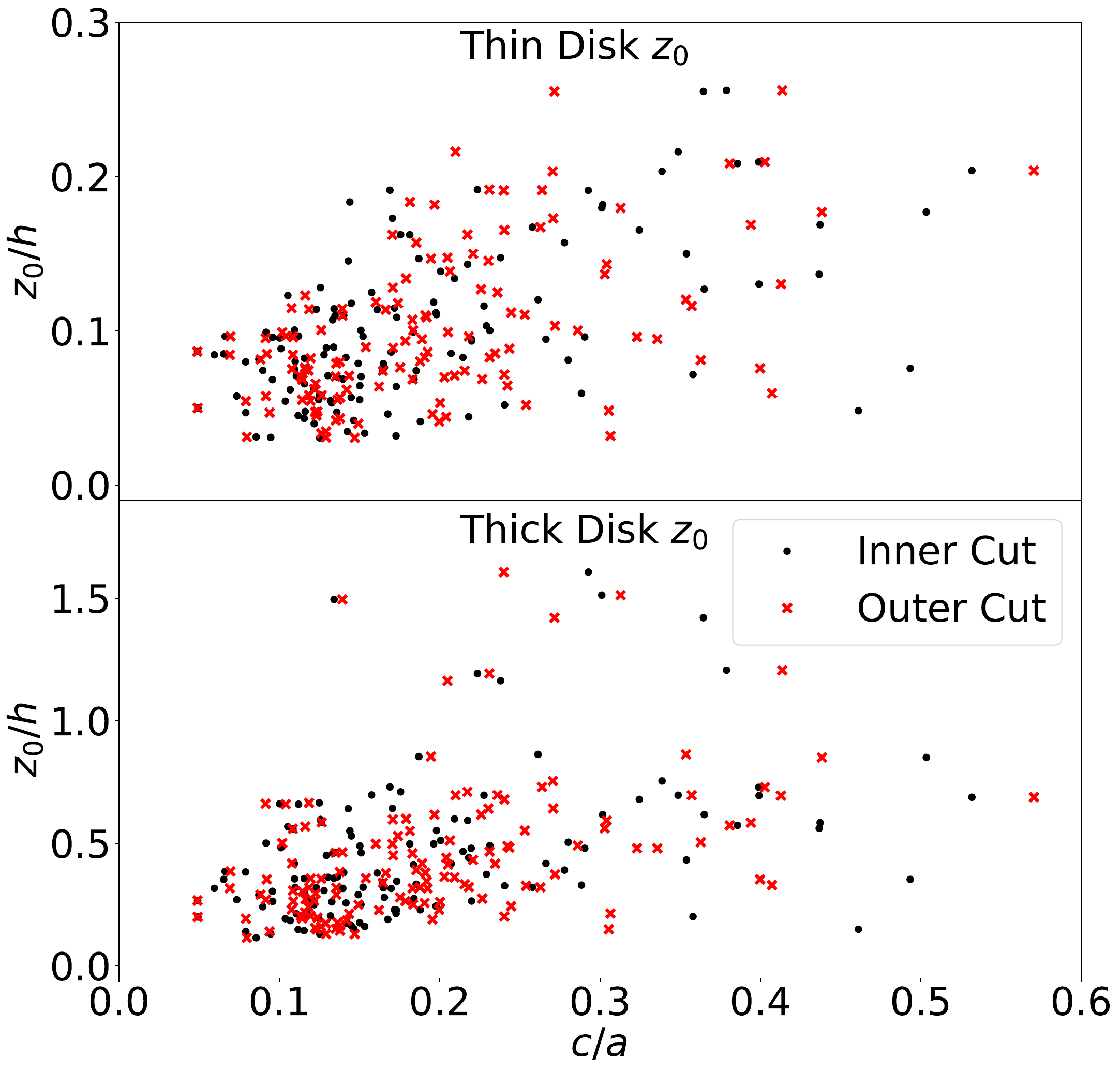}
    \caption{
    Ratio of scale height to scale length plotted against intrinsic flattening, $c/a$.
    Top/Bottom panel: thin/thick disk scale height.
    Black/red dots indicate intrinsic flattening values derived from inner/outer cuts.
    Note the different Y-axis limits in both figures.}
    \label{fig:z/h_vs_c/a}
\end{figure}

\begin{figure}
    \centering
    \includegraphics[width=\linewidth]{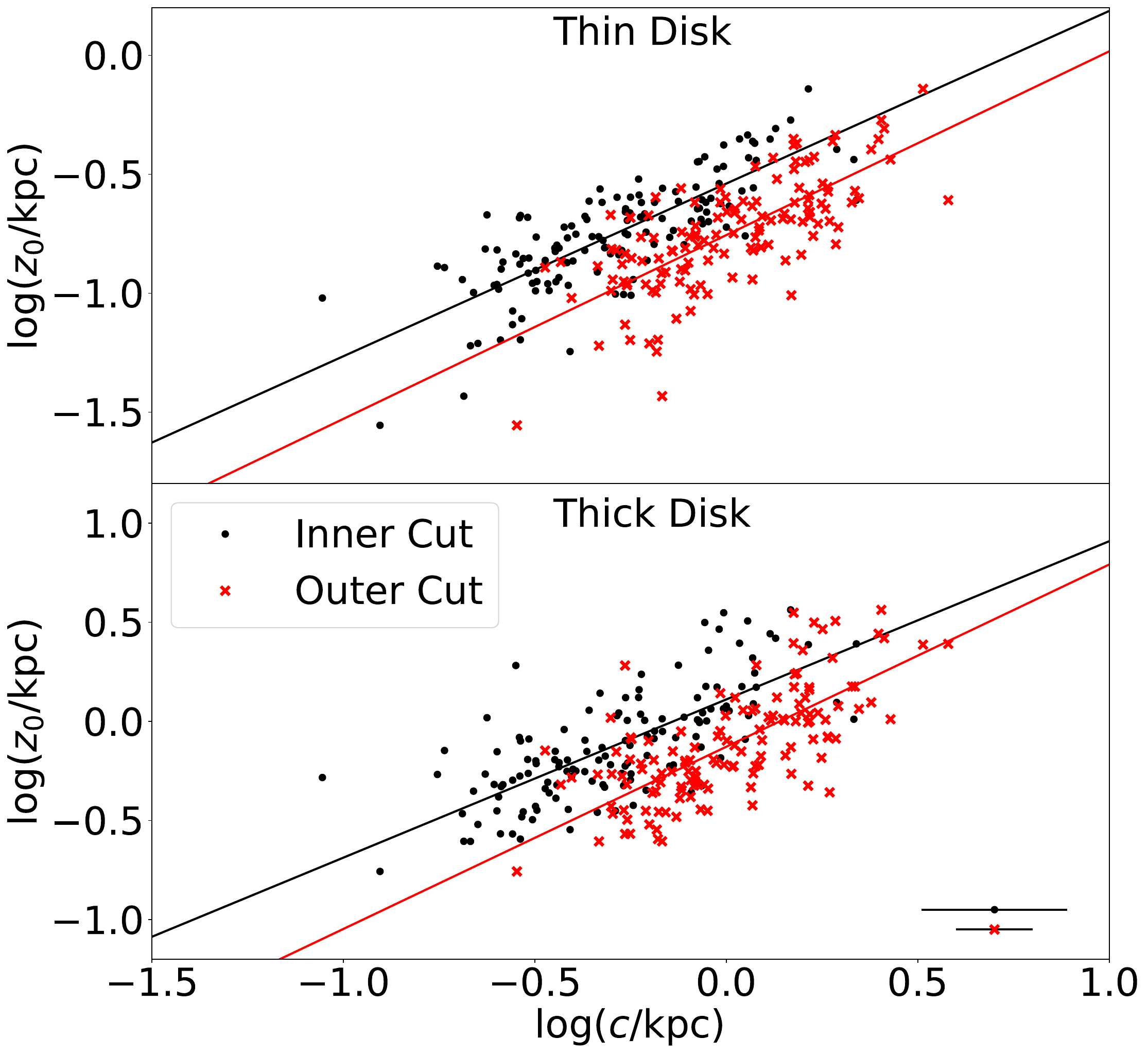}
    \caption{Disk scale height versus characteristic cut's semi-minor axis.
    Top/bottom panels show thin/thick disk scale height.
    Black/red dots indicate intrinsic flattening values derived from inner/outer cuts. 
    The solid black/red line is a linear fit to these data.}
    \label{fig:z_vs_c}
\end{figure}

\begin{figure}
    \centering
    \includegraphics[width=\linewidth]{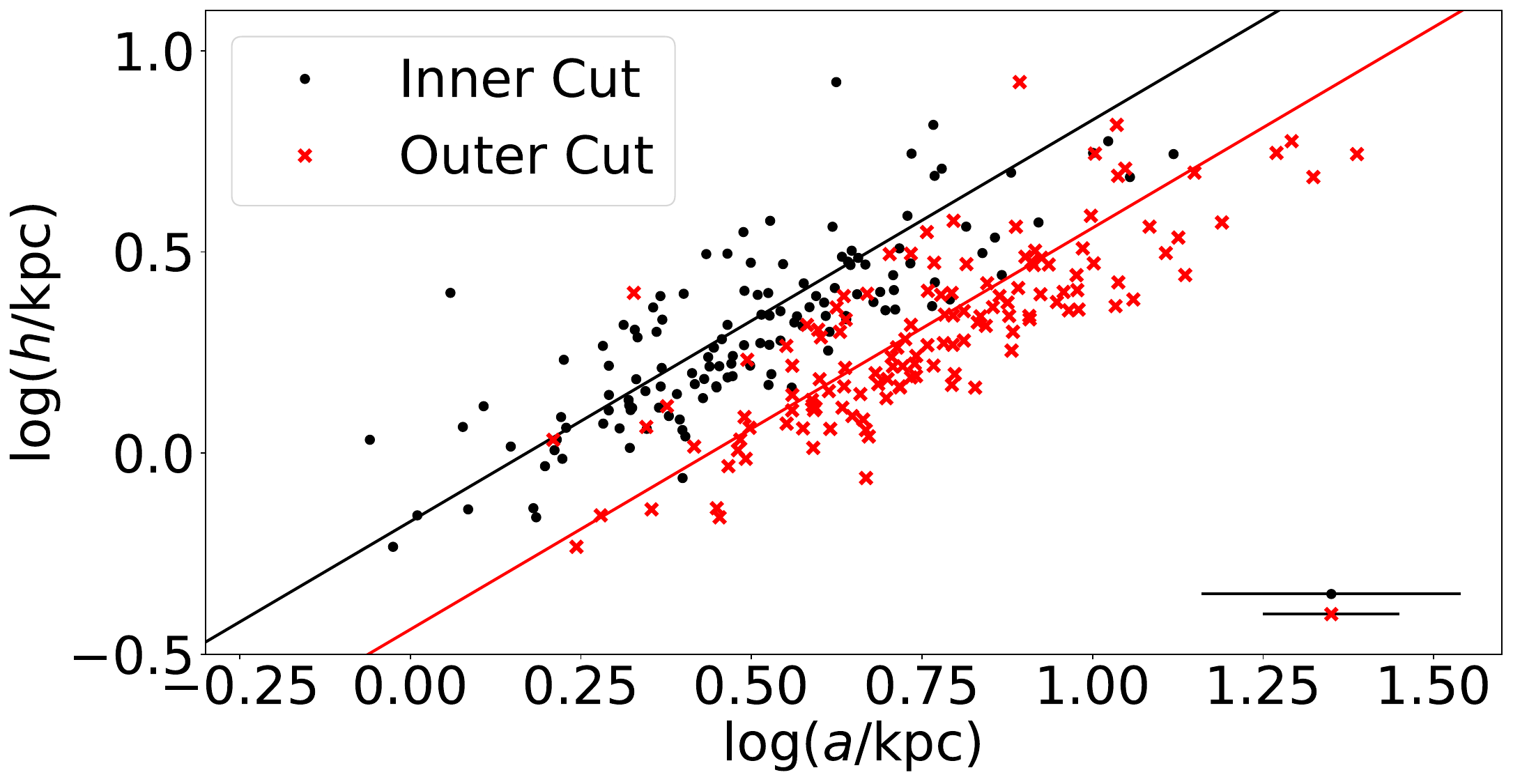}
    \caption{Disk scale length versus characteristic cut's semi-major axis.
    The scale lengths represent the global disk, not a thin/thick disk decomposition.
    Black/red dots represent intrinsic flattening values derived from inner/outer cuts.
    The solid black/red line is a linear fit to these data.}
    \label{fig:h_vs_a}
\end{figure}

Parametric decompositions of galaxies have become ubiquitous in studies of vertical disk structure \citep[e.g.,][]{Gilmore1983, Tikhonov2005, Yoachim2006, Comeron2011a, Comeron2011b, Comeron2011c, Comeron2012, Erwin2015, Comeron2018}.
We now test whether parametric and non-parametric measures of disk thickness map into one another.
As stated above, relevant model parameters include scale heights, $z_0$, of the thin and thick disks, here taken from the SB profile decompositions of \citet{Comeron2018}.
We have also extracted a global scale length from our own exponential fits to the disk-dominated SB profiles of our S\textsuperscript{4}G galaxies, since the scale lengths of \citet{Comeron2018} may correspond to non-disk structures. 
With these values, we can test for any correlations between $c/a$ and $z_0/h$ (\Fig{z/h_vs_c/a}), as well as between $h$ and $a$ (\Fig{h_vs_a}), and $z_0$ and $c$ (\Fig{z_vs_c}).

To investigate these trends, we have computed the average value of $a$ for both inner and outer cuts as $a_{\rm{in}} = 0.35\,{r_{25}}$ and $ a_{\rm{out}} = 0.65\,{r_{25}}$, respectively.
To estimate the error on these values, we assigned both values of $a$ the maximum possible uncertainty of $a_{\rm{err}} = 0.15\,{r_{25}}$, which is the distance from the center to the edge of either cut.
As a result of this choice, the percent errors on $a$ will be identical for a given cut: $42\%$ error for the inner cut and $23\%$ error for the outer cut.
It follows that the semi-minor axis of the characteristic cut is
$c_i = a_i(c/a)_i\mathrm{,}$ where $i$ denotes either the inner or outer cut.
The uncertainty on $c/a$ and $a$ is propagated through to $c$.
With values of $a$ in place, values of $c$ emerge directly from the ratio $c/a$.

Weak correlations between $z_0/h$ and $c/a$, as supported by the Pearson correlation coefficients reported in \Table{scale_fit_params}, are found in \Fig{z/h_vs_c/a}.
Somewhat stronger correlations ($r_p \geq 0.7$) exist between $z_0$ and $c$ as well as $h$ and $a$, as seen in \Fig{z_vs_c} and \Fig{h_vs_a}, respectively.
These suggest that our method and choice of cuts are indeed probing disk structure and properly avoiding the CMC.

The thin disk correlations in \Table{scale_fit_params} are weaker for the thick disk, likely as a result of a combination of lower S/N in the fainter regions and heightened sensitivity to perturbations in the outskirts.
The $\log{h} \! - \! \log{a}$ relation is an exception, as the definition of $a$ and choice of global $h$ force identical scatters between the inner and outer cuts.

\begin{table}
    \centering
    \begin{tabular}{c|cccc}
        \toprule
          & \multicolumn{4}{c}{$z_0/h = m(c/a) + b$}\\
          \cmidrule(lr){2-5}
          & \multicolumn{2}{c}{Thin Disk} & \multicolumn{2}{c}{Thick Disk} \\
          \cmidrule(lr){2-3}\cmidrule(lr){4-5}
          & Inner Cut & Outer Cut & Inner Cut & Outer Cut \\
          \cmidrule(lr){2-2}\cmidrule(lr){3-3}\cmidrule(lr){4-4}\cmidrule(lr){5-5}
         $m$ & $0.28$ & $0.28$ & $1.2$ & $1.3$ \\
         $\Delta m$ & $0.03$ & $0.04$ & $0.2$ & $0.2$ \\
         $b$ & $0.05$ & $0.05$ & $0.23$ & $0.19$ \\
         $\Delta b$ & $0.01$ & $0.01$ & $0.05$ & $0.05$ \\
         $r_p$ & $0.57$ & $0.52$ & $0.42$ & $0.42$ \\
         $p$ & $\ll 1\%$ & $\ll 1\%$ & $\ll 1\%$ & $\ll 1\%$ \\
        \midrule
          & \multicolumn{4}{c}{$\log{z_0} = m\log{c} + b$}\\
          \cmidrule(lr){2-5}
          & \multicolumn{2}{c}{Thin Disk} & \multicolumn{2}{c}{Thick Disk} \\
          \cmidrule(lr){2-3}\cmidrule(lr){4-5}
          & Inner Cut & Outer Cut & Inner Cut & Outer Cut \\
          \cmidrule(lr){2-2}\cmidrule(lr){3-3}\cmidrule(lr){4-4}\cmidrule(lr){5-5}
         $m$ & $0.72$ & $0.77$ & $0.80$ & $0.92$ \\
         $\Delta m$ & $0.05$ & $0.06$ & $0.07$ & $0.08$ \\
         $b$ & $-0.54$ & $-0.76$ & $0.11$ & $-0.13$ \\
         $\Delta b$ & $0.02$ & $0.01$ & $0.03$ & $0.02$ \\
         $r_p$ & $0.78$ & $0.71$ & $0.71$ & $0.70$ \\
         $p$ & $\ll 1\%$ & $\ll 1\%$ & $\ll 1\%$ & $\ll 1\%$ \\
        \midrule
          & \multicolumn{4}{c}{$\log{h} = m\log{a} + b$}\\
          \cmidrule(lr){2-5}
          & \multicolumn{2}{c}{Inner Cut} & \multicolumn{2}{c}{Outer Cut} \\
          \cmidrule(lr){2-3}\cmidrule(lr){4-5}
          $m$ & \multicolumn{2}{c}{$1.00$} & \multicolumn{2}{c}{$1.00$} \\
          $\Delta m$ & \multicolumn{2}{c}{$0.06$} & \multicolumn{2}{c}{$0.06$} \\
          $b$ & \multicolumn{2}{c}{$-0.17$} & \multicolumn{2}{c}{$-0.44$} \\
          $\Delta b$ & \multicolumn{2}{c}{$0.03$} & \multicolumn{2}{c}{$0.04$} \\
          $r_p$ & \multicolumn{2}{c}{$0.81$} & \multicolumn{2}{c}{$0.81$} \\
          $p$ & \multicolumn{2}{c}{$\ll 1\%$} & \multicolumn{2}{c}{$\ll 1\%$} \\
        \bottomrule
        
    \end{tabular}
    \caption{The fit parameters for the linear functions $\frac{z_0}{h}(c/a)$ (top third), and power laws $z_0(c)$ (middle third), and $h(a)$ (bottom third).
    Fit parameters are the slope $m$ and y-intercept $b$, along with their respective errors $\Delta m$ and $\Delta b$.
    $r_p$ is the Pearson correlation coefficient of the data.
    $p$ is the two-sided p-value that denotes the significance of the correlation.}
    \label{tab:scale_fit_params}
\end{table}

Overall, while some tentative trends ($r_p \geq 0.7$) between $z_0$ and $c$, as well as between $h$ and $a$, exist and are indeed expected on the basis of simple galactic models, a more complex, likely multi-dimensional, analysis of all the galaxy structural variables at play would be required to achieve a closer mapping between parametric and non-parametric tracers of disk thickness.
As we shall see in \sec{scale_v_isophote}, the ratio $c/a$ remains a better tracer of galaxy disk thickness than $z_0/h$.

\begin{figure}
    \centering
    \includegraphics[width=\linewidth]{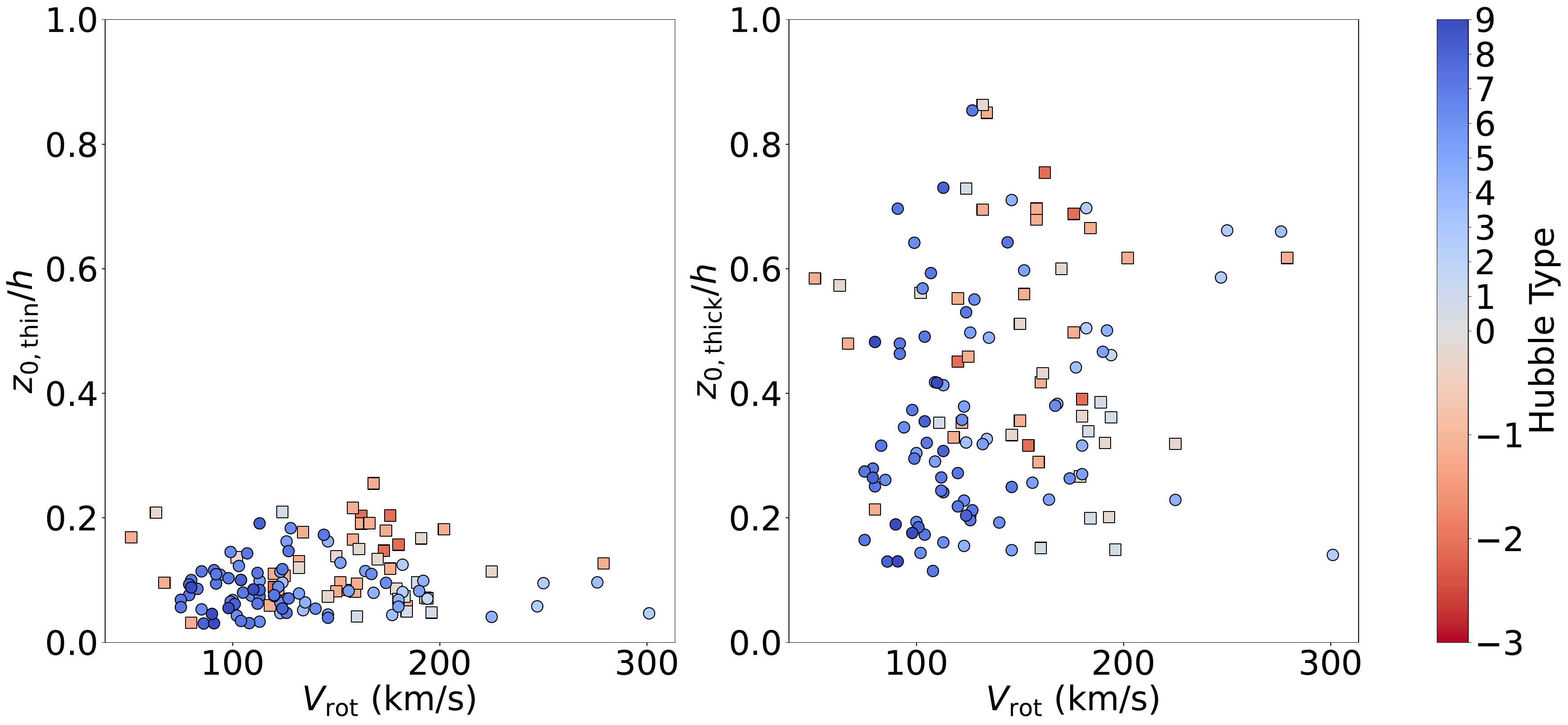}
    \caption{Measure of parametric flattening for the thin (left panel) and thick (right panel) disks of spiral and lenticular galaxies versus total rotational speed, $V_{\rm{rot}}$. 
    In both cases, no correlation between $z_{0,\rm{thin}}/h$ and $V_{\rm{rot}}$ is found.
    For the thin disks of spiral galaxies (blue circles in the left panel), the mean $z_{0,\rm{thin}}/h = 0.12$ is shown as a thick line.
    Symbols and colours are as in \Fig{other_params}.
    }
    \label{fig:scale_ratio_vrot}
\end{figure}

We close this sub-section by revisiting the (non-)correlation of flattening parameter with $V_{\rm{rot}}$ seen in \sec{q+other} (\Fig{other_params}, panel (i)). 
The latter is surprising as one would expect the depth of the potential and magnitude of rotational support to affect the stellar disk thickness \citep[e.g.,][]{deGrijs1997, Kregel2002, Pohlen2004}, as it does with dusty disks \citep[e.g.,][]{Dalcanton2004,Mosenkov2022}. 
We can here revisit the $c/a$ vs $V_{\rm{rot}}$ relation in terms of the flattening parametric $z_0/h$ in order to compare with the references above. 
This is shown in \Fig{scale_ratio_vrot}.
Values of $z_0$ for the thin and thick disks are still from \citet{Comeron2018}.
While the data in \citet{Kregel2002} agree with ours (\Fig{scale_ratio_vrot}), our interpretations for a putative trend between $z/h$ (or in our case $c/a$) and $V_{\rm{rot}}$ differ, owing in part to our larger data base.
Our tests with bona fide $z_0/h$, or $c/a$, versus $V_{\rm{rot}}$ yield the same conclusion: there is no dependence of $V_{\rm{rot}}$ on disk flattening. 

\subsection{Galactic Environments}
\label{sec:environments}

\begin{figure}
    \centering
    \includegraphics[width=\linewidth]{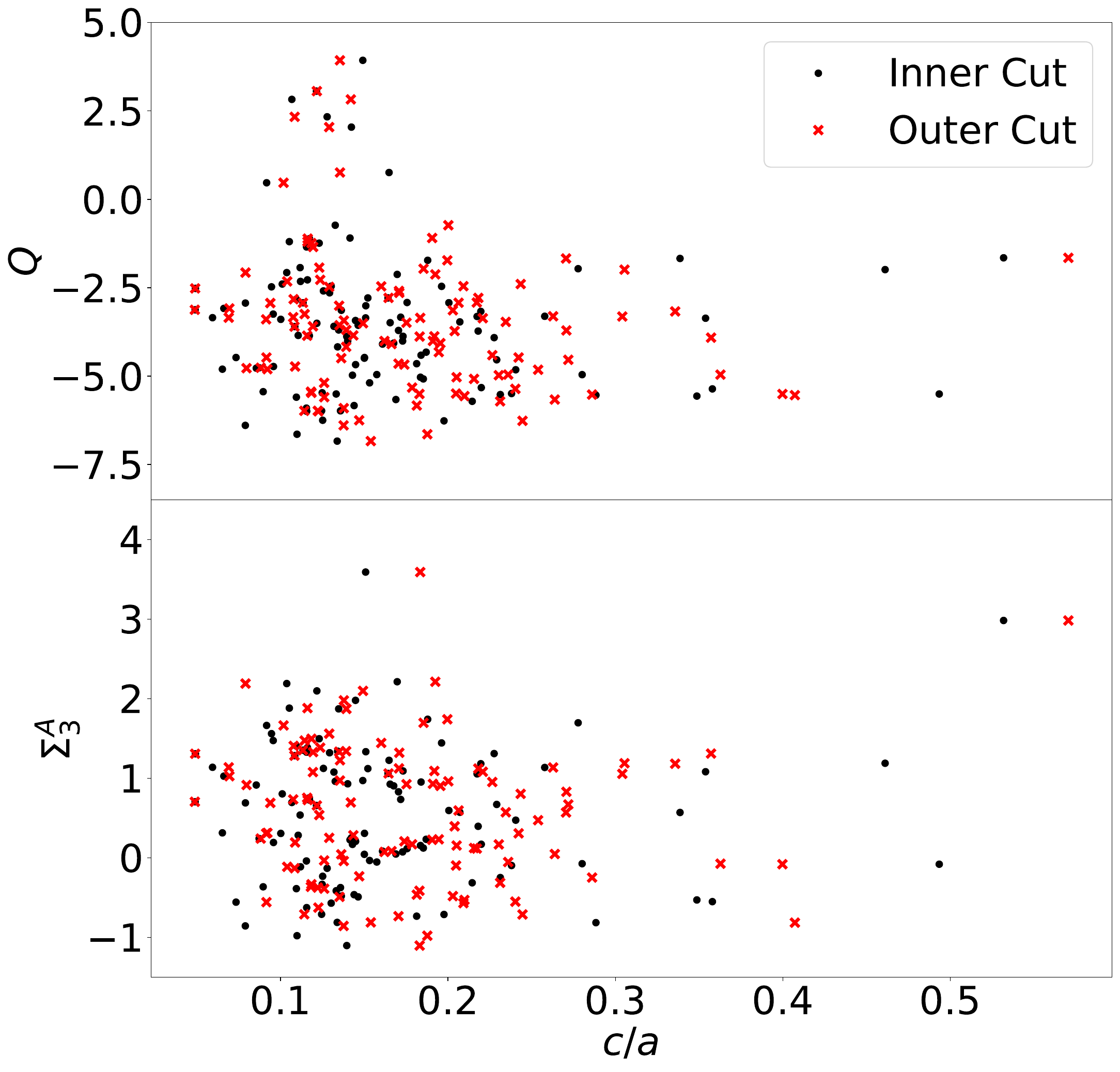}
    \caption{Top: Dahari parameter of each galaxy against its inner (blue) and outer (black) intrinsic flattening.
    Bottom: projected surface density of galaxies to the third nearest neighbour for each galaxy against its inner (blue) and outer (black) intrinsic flattening.}
    \label{fig:environment_vs_flattening}
\end{figure}

The shapes of S\textsuperscript{4}G galaxies could be affected by environmental effects, such as tidal interactions from neighbour galaxies.
We used the environmental quantifiers described in \sec{Pparam} to test for correlations between these parameters and intrinsic flattening.
Any correlation could suggest whether galactic environments play a key role in driving disk thickness, which would need to be accounted for when deriving the intrinsic $c/a$.

However, the top panel of \Fig{environment_vs_flattening} shows that the Dahari parameter is not correlated with intrinsic flattening (see \Table{environment_pearson_r}.)
Although some notably large $Q$ values indicate the presence of significant tidal forces, the stellar disks of the galaxies experiencing such forces seem neither unusually thick nor thin.

\begin{table}
    \centering
    \begin{tabular}{c|cccc}
        \toprule
          & \multicolumn{2}{c}{Inner Cut} & \multicolumn{2}{c}{Outer Cut} \\
          \cmidrule(lr){2-3}\cmidrule(lr){4-5}
          & $r_p$ & $p$ & $r_p$ & $p$ \\
          \cmidrule(lr){2-2}\cmidrule(lr){3-3}\cmidrule(lr){4-4}\cmidrule(lr){5-5}
         $Q$ & $-0.09$ & $\mathbf{36\%}$ & $-0.16$ & $\mathbf{9\%}$ \\
         $\Sigma^{A}_{3}$ & \phantom{$-$}$0.03$ & $\mathbf{72\%}$ & $-0.01$ & $\mathbf{95\%}$ \\
        \bottomrule
        
    \end{tabular}
    \caption{Pearson correlation coefficients and their significance for the correlation between environment tracers plotted in \Fig{environment_vs_flattening} and intrinsic flattening.
    $r_p$ is the Pearson correlation coefficient.
    $p$ is the two-sided p-value that denotes the significance of the correlation.
    Boldface p-values are those that fail the significance test.}
    \label{tab:environment_pearson_r}
\end{table}

The bottom panel of \Fig{environment_vs_flattening} shows an equally large scatter.
As reported in \Table{environment_pearson_r}, the surface density of the galaxy group hosting the subject does not correlate with intrinsic flattening.
There are two outlying galaxies of interest with large group surface densities.
The first, at $0.1 \leq c/a \leq 0.2$, does not boast an abnormal intrinsic flattening.
Therefore, we do not expect this galaxy's disk to be biased by its environment.
The second outlier lies at $0.5 \leq c/a \leq 0.6$ in both cuts.
The dense environment of this galaxy may be causing the disk to thicken.

Overall, the lack of correlation between our environmental quantifiers suggests that our sample is not affected by close interactions between galaxies.

\section{Discussion}
\label{sec:Discussion}

In \sec{Flattening}, we computed and reported several measurements of $c/a$ for spiral and lenticular galaxies in \Table{gaussian_params}.
We further subdivided these samples and reported the median $c/a$ of each morphological type in \Table{flattening_type_medians}. 
In what follows, we identify the most reliable $c/a$ based on our method of characteristic cuts.
We further cement these results with a comparative analysis of the parametric $z_0/h$ in \sec{scale_v_isophote}.

\subsection{The Intrinsic Flattening of Stellar Disks}\label{sec:prescribe_flattening}

We presented in \sec{q_distribution} three distinct measures of $c/a$ for spiral and lenticular galaxies, each applied to the inner and outer cuts (both representing the thin and thick disks) of our galaxy sample.
Those measurements of $c/a$ were: (i) the Gaussian mean of the $c/a$ distribution (see \Fig{c/a_hist_fit}); (ii) 
the median $c/a$ for each Hubble Type, tabulated in \Table{flattening_type_medians} and plotted in \Fig{flattening_by_T}; and (iii) the average of these median $c/a$ values for spiral morphologies, visualized as the horizontal lines in \Fig{flattening_by_T}.
The variance in \Fig{c/a_hist_fit} and \Fig{flattening_by_T} made clear that a global $c/a$ cannot be reasonably ascribed to lenticular galaxies.
Even when binned by Hubble Type, the large scatter persists.
It follows that a single estimate for the intrinsic flattening of lenticular systems is not realistic.

For disk galaxies, the tight scatter of the distributions of \Fig{c/a_hist_fit} makes the results from our Gaussian fits more compelling than the median $c/a$ values for each spiral morphology in \Table{flattening_type_medians} due to the larger uncertainties and under-sampled bins of the latter.
Line fitting these median values would add needless complexity and error.

There is little statistical difference between the inner and outer cut Gaussian fits to the spiral distributions reported in \Table{gaussian_params}, the inner cut having a marginally tighter scatter.
The tail in the outer cut's distribution (see \Fig{c/a_hist_fit}) also takes that distribution away from a true Gaussian.
Given the tight scatter and more Gaussian distribution for the inner cut, we take the result $\langle c/a \rangle = 0.124 \pm 0.001~(\rm{stat}) \pm 0.033~(\rm{intrinsic/systematic})$ as most representative for the intrinsic flatness of spiral galaxies.

Our result agrees well with the non-parametric studies (see \sec{history}) of \citet{Guthrie1992}, based on micrometer measurements, and \citet{Giovanelli1994}, who did not provide uncertainties.

\subsection{The Effectiveness of the Method of Characteristic Cuts}
\label{sec:test_isophotes}

\begin{figure}
    \centering
    \includegraphics[width=\textwidth,height=\textheight,keepaspectratio]{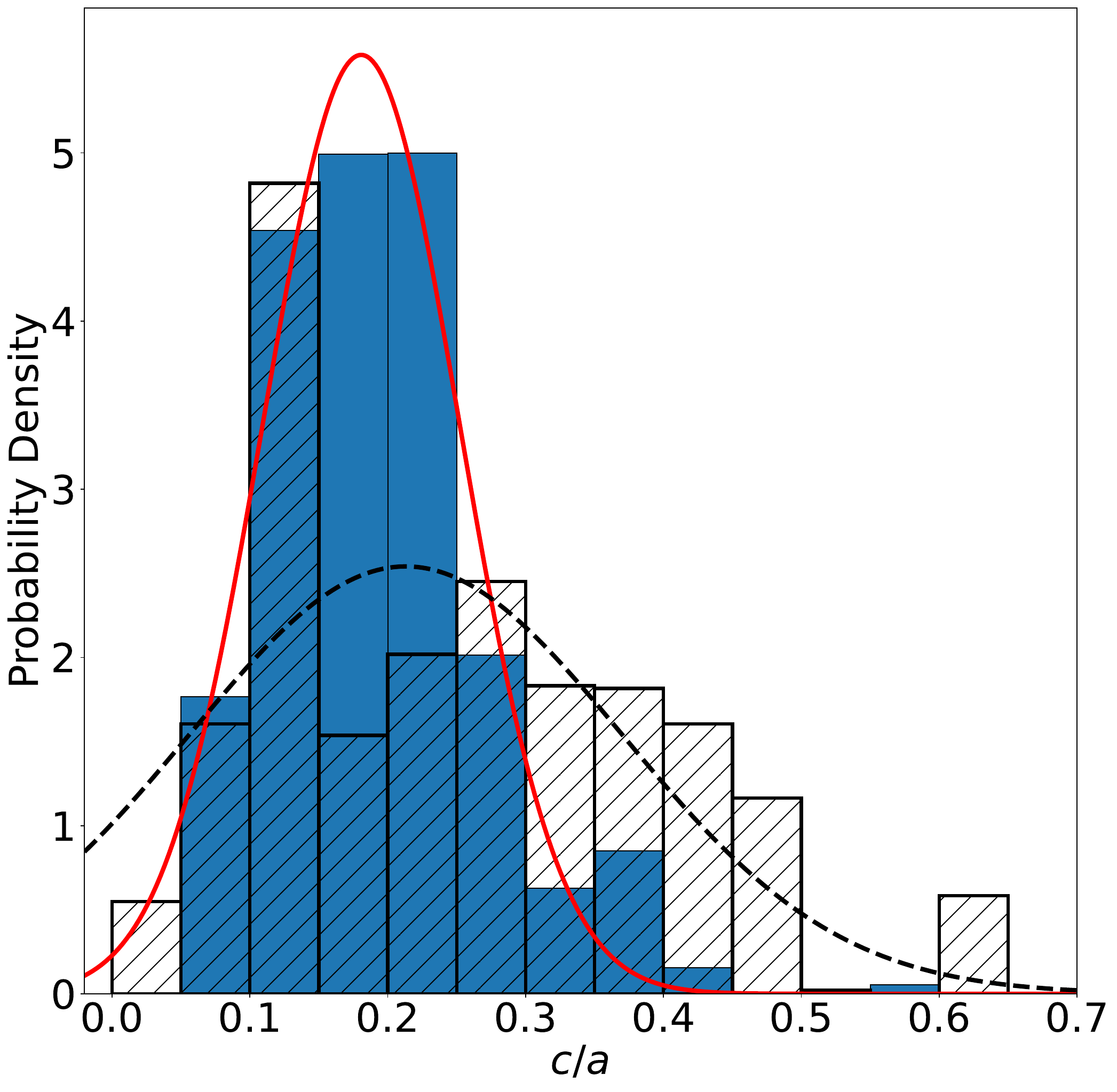}
    \caption{Distribution of intrinsic flattening measured at $r = a = {r_{25}}$, for the spiral and lenticular populations.
    The solid blue and hatched histograms represent the distribution of spiral and lenticular galaxies, respectively. 
    The solid red and dashed lines are Gaussian fits to the spiral and lenticular distributions with $\mu(Sp) = 0.175 \pm 0.003$ and $\sigma = 0.071 \pm 0.002$ and $\mu(S0) = 0.21 \pm 0.03$ and $\sigma = 0.16 \pm 0.02$, respectively.
    }
    \label{fig:r_25_q_hist}
\end{figure}

To assess whether the method of characteristic cuts produces the most accurate measurements of intrinsic flattening, we compared our results to those derived from the more straightforward method of choosing a singular isophote.
A traditional isophotal reference is $\mu_B = 25$ \magss.
\Fig{r_25_q_hist} shows the error-weighted distribution of our sample spiral and lenticular's intrinsic flattening measured from the isophote of radius $r = {r_{25}}$, or whichever isophote whose radius was closest to ${r_{25}}$.

Comparing the properties of this distribution (see \Fig{r_25_q_hist}'s caption) with those derived from our method of characteristic cuts (see \Table{gaussian_params}), our method, especially as applied to the inner cut, is found to minimize uncertainties and yield lower variance in $c/a$ for both the spiral and lenticular galaxies.
In short, a singular isophote traces heterogeneous structure resulting in a poorer tracer of intrinsic disk thickness.
Furthermore, outer isophotes are more susceptible to lower S/N and perturbations.
Our method based on characteristic cuts avoids these issues.

We have further compared the correlations between the structural parameters explored in \sec{q+other} with those in \Table{other_params_pearson_r}.
We tested for correlations between $c/a|_{r_{25}}$ and these parameters.
Cross-referencing the resulting correlations against \Table{other_params_pearson_r} showed no significant changes.
We conclude that our method of characteristic cuts yields a $c/a$ distribution that is tighter than that of previous techniques, including \sec{scale_v_isophote} below. 

\subsection{Parametric Decompositions as Measures of Intrinsic Flattening}
\label{sec:scale_v_isophote}

In principle, the ratio $z_0/h$ can also be used to assess the global flattening of galaxy disks, provided the disk structures are well described by the adopted models.
In \sec{ComeronCompare}, we already assessed that neither the thin nor thick disk $z_0/h$ map closely into $c/a$.
However, the small range of $z_0/h$ values for the thin disk offers a valuable comparison with our $c/a$ values. 

\begin{figure}
    \centering
    \includegraphics[width=\textwidth,height=\textheight,keepaspectratio]{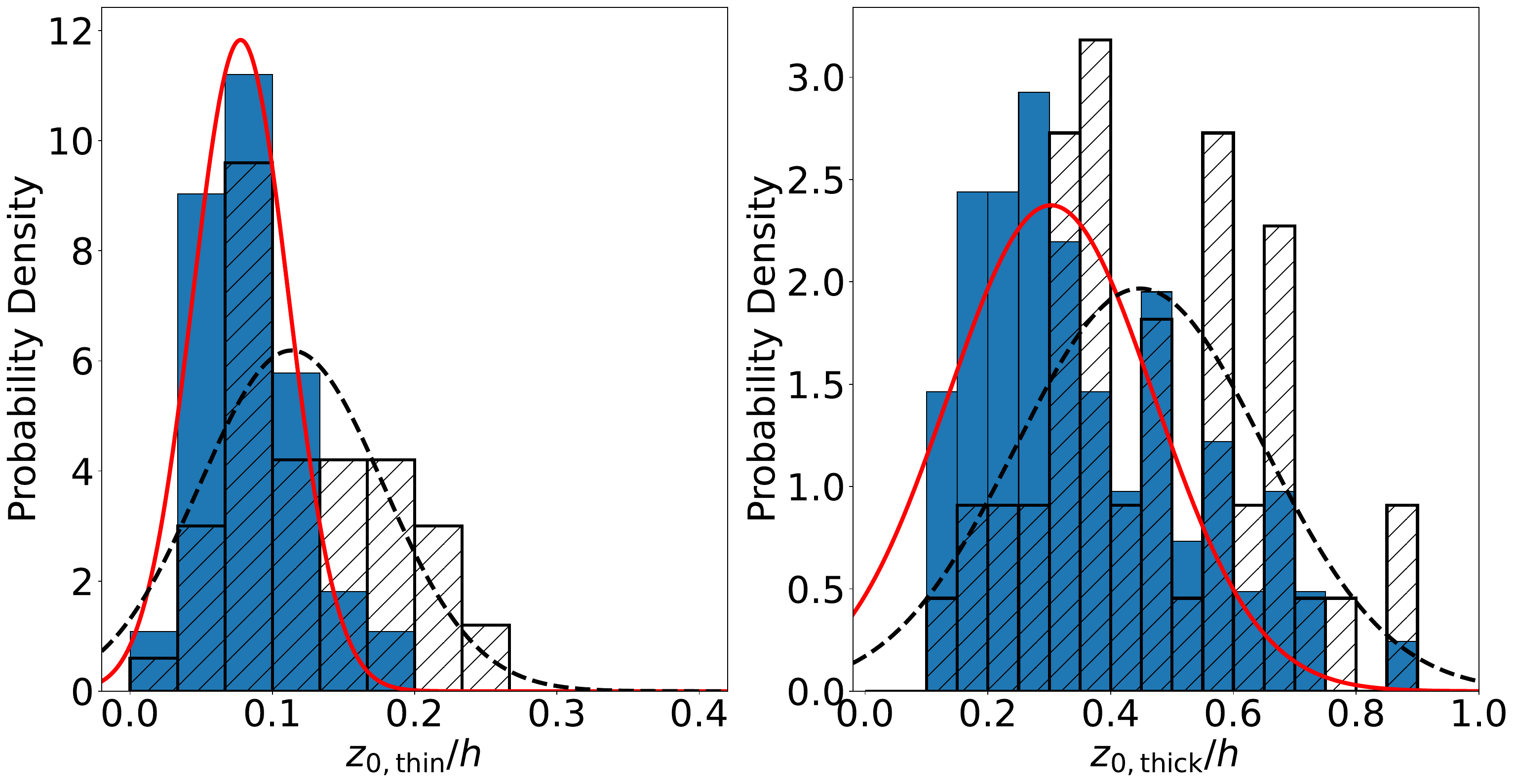}
    \caption{Histograms of the distributions of $z_0/h$ for thin (left) and thick (right) disks. 
    The solid blue histograms are the distributions of spiral galaxy $z_0/h$, and the hatched histograms are the distributions of lenticular galaxy $z_0/h$.
    Left: The solid red line is a Gaussian fit to the spiral distribution with $\mu = 0.078 \pm 0.001$ and $\sigma = 0.034 \pm 0.001$.
    The dashed line is a Gaussian fit to the lenticular distribution with $\mu = 0.113 \pm 0.008$ and $\sigma = 0.064 \pm 0.007$.
    Right: The solid red line is a Gaussian fit to the spiral distribution with $\mu = 0.30 \pm 0.02$ and $\sigma = 0.17 \pm 0.02$.
    The dashed line is a Gaussian fit to the lenticular distribution with $\mu = 0.45 \pm 0.04$ and $\sigma = 0.20 \pm 0.03$.
    }
    \label{fig:scale-ratio-hists}
\end{figure}

\Fig{scale-ratio-hists} shows the distributions of $z_0/h$ values for the thin and thick disk scale heights.
Our final estimate for the intrinsic $c/a$ ought to lie somewhere between the values of $z_0/h$ for the thin and thick disk, as is indeed found here. 
Gaussian fits to these distributions for spiral galaxies yield a mean flattening ratio for the thin disk of $\langle z_{0,\rm{thin}}/h \rangle \approx 0.08$ and thick disk $\langle z_{0,\rm{thick}}/h \rangle \approx 0.30$. 
These are viewed as lower and upper limits for the intrinsic flattening of galaxy disks. 
In a multiple disk system, the intrinsic flattening might roughly correspond to the superposition of thin and thick disk parameters, so in a galaxy with a thin/thick disk dichotomy, $c/a$ should be greater than $z_{0,\rm{thin}}/h$.
The thin disk value is slightly smaller than our reported mean $c/a$, providing some assurance that our method of characteristic cuts truly measures intrinsic flattening.
This exercise suggests that we have converged on a rather accurate estimate for the intrinsic flattening of spiral galaxies.

The Gaussian fits also show broader distributions of $z_0/h$, with a superposition of $\sigma_{\rm{thin}} = 0.03$ and $\sigma_{\rm{thick}} = 0.17$, while $c/a$ is measured with a single deviation of 0.03.
Once again this supports using characteristic cuts to measure of the disk intrinsic thickness since it produces scatter comparable to the parametric $z_0/h$ via a simpler method.

Furthermore, our $\langle c/a \rangle$ value agrees well with the $\langle z_0/h \rangle$ reported by each of \citet{Kregel2002}, \citet{Hernandez2006}, \citet{Mosenkov2015}, and \citet{Mosenkov2022}.
That we arrive to the same intrinsic flattening with a simpler model-independent technique highlights the value and benefits of our method.

\section{Future Work}\label{sec:Future_Work}

As analytical methods evolve, so does the quality of datasets. 
The proposed method of characteristic cuts to constrain the intrinsic flattening of galaxy disks will certainly benefit from greater sample sizes. 
A ten-fold increase in the size of our sample, especially for the lenticulars, would go a long way towards characterizing the changes of disk thickening with Hubble type. 

Considering the current data set, the S\textsuperscript{4}G images were reduced over a series of five pipelines \citep{Sheth2010, MunosMateos2015, Salo2015, Querejeta2015} to produce a variety of science-ready data products.
In this study we analyzed maps from Pipeline 1 and used masks from Pipeline 2, but additional products, such as independent measurements of axial ratios and stellar emission maps, may prove valuable to further validate our findings.
For instance, the S\textsuperscript{4}G Pipeline 5 \citep{Querejeta2015} separated galaxy images into stellar and dust emission maps using the Independent Component Analysis method of \citet{Meidt2012}.
The effects of dust on the characteristic cuts used in this study could be assessed by repeating our analysis using these stellar emission maps, thereby quantifying the flattening of isophotes caused by dust emission.
However, the use of such maps might introduce model-dependences that we aimed to avoid.

The foundation for the estimation of disk intrinsic flattening from face-on view \citep[e.g.,][]{Giovanelli1994} is an ad-hoc formulation that assumes the observed flux varies linearly with $a_{\rm{obs}}/c_{\rm{obs}}$.
Empirical measurements of inclination effects cannot be made via direct observations.
Instead, galaxy simulations could be observed to characterize individual systems from multiple angles.
For instance, the Numerical Investigation of a Hundred Astrophysical Objects project \citep[NIHAO;][]{Wang2015} provides impressive access to a wide variety of simulated galaxy types.
Mock observations of NIHAO disk galaxies at various inclination angles, primarily focused around near face- and edge-on perspectives, offer an ideal data set with which to (i) repeat this analysis and verify our results, and (ii) derive an empirical formulation of the inclination effects on isophotal radii and observed axis ratios 
\citep[e.g.,][]{Byun1994,Agertz2013,Roskar2014}.

\section{Summary}\label{sec:Conclusion}

Defining the size of a galaxy based on its projected image on the sky is especially challenging and requires care.
We have proposed the method of characteristic cuts for measuring the intrinsic flattening of disk galaxies and applied it to a sample of 141 (nearly) edge-on galaxies from the S\textsuperscript{4}G imaged in $3.6 \ \mu$m.
We used AutoProf, a Python-based  automated tool for the photometry of astronomical images
\citep{Stone2021b}, to perform isophotal fits on the $3.6 \ \mu$m images of each galaxy. 
With these isophotes, we established the characteristic cuts of each galaxy by taking the average axial ratio $c/a$ of isophotes falling within axial cuts of $0.2\,{r_{25}} \leq |a| \leq 0.5\,{r_{25}}$ and $0.5\,{r_{25}} \leq |a| \leq 0.8\,{r_{25}}$, which we referred to as the inner and outer cuts, respectively.
Those galaxies which showed clear visual indications of poor fits or had large $c/a$ variances within either cut were rejected, leaving us with a sample of 133 galaxies with good fits.
We also retrieved surface brightness profiles and thin and thick disks decompostions from \citet{Comeron2018} in addition to various physical parameters from \citet{MunosMateos2015}.
Our findings are summarized here:

\begin{enumerate}
\item Spiral galaxies follow a universal intrinsic disk flattening with a mean value of  
 $\langle c/a \rangle = 0.124 \pm 0.002$ and standard deviation $\sigma = 0.034 \pm 0.001$. 

\item For lenticular galaxies, the $c/a$ distribution is broader, with standard deviation $\sigma = 0.13 \pm 0.02$ and, after averaging the inner and outer cuts, is centered on $\langle c/a \rangle = 0.20 \pm 0.04$.
For $T < 3$, $c/a$ negatively correlates with $T$, whereas $c/a$ is independent of $T$ for $T > 3$.
Lenticular galaxies show a conspicuously strong negative correlation with Hubble type, though the scatter in $c/a$ is too large to make definitive claims about the evolution of their disks.
Galaxies with bars do not appear significantly thicker than those without, though barred systems in our sample are too few to claim that this is a true meaningful difference.

\item While largely insensitive to any galaxy structural parameter, including circular velocity, the intrinsic flattening of galaxy disks shows a tentative correlation with $\log{C_{31}}$, in the sense that bulge and disk growth are linked
\citep{Courteau1996,Kormendy2004, Comeron2014}.

\item The scale height, $z$, correlates well with the vertical size $c$, and the scale length, $h$, correlates well with the semi-major axis length, $a$, but the ration of $z/h$ does not correlate well with $c/a$. 

\end{enumerate}


\section{Acknowledgments}

This research project was initiated under the watchful eye of Simon D{\'i}az Garc{\'i}a.
We are grateful for his precious suggestions and contributions. 
Peter Erwin is also thanked for insightful discussions and references.
SC(Canada) and CS acknowledge generous support from the Natural Sciences and Engineering Research Council of Canada and Queen's University through various scholarships and grants. 
CS also acknowledges support from the Canadian Institute for Theoretical Astrophysics (CITA) National Fellowship program.
SC(Spain) also acknowledges funding from the State Research Agency (AEI-MCINN) of the Spanish Ministry of Science and Innovation under the grant ``Thick discs, relics of the infancy of galaxies'' with reference PID2020-113213GA-I00.

\appendix
\section{Resolution Effects and PSF} \label{sec:AppendixA}

\begin{figure}
    \centering
    \includegraphics[width=\linewidth]{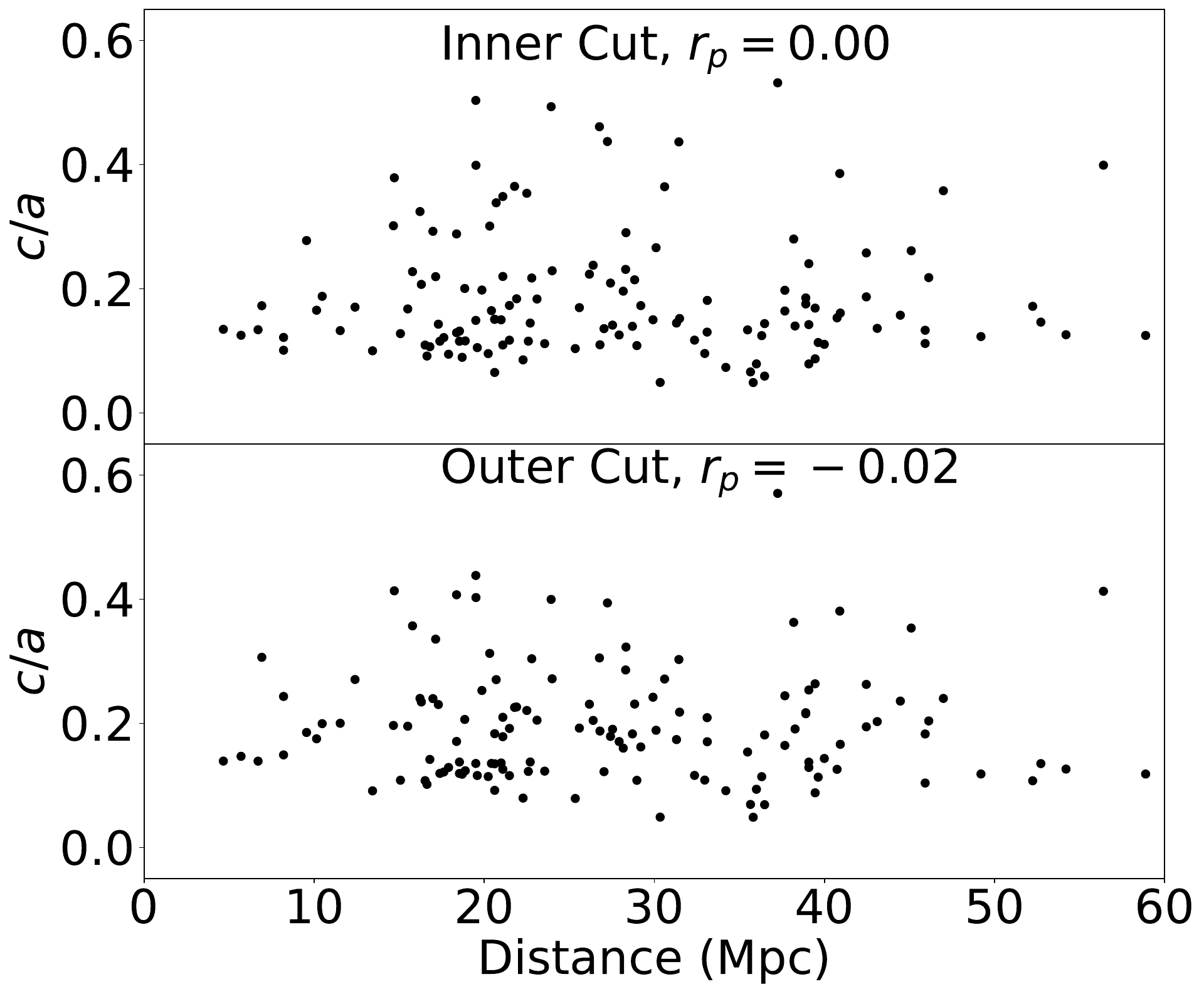}
    \caption{Intrinsic flattening of the inner (top) and outer (bottom) cuts as a function of distance.}
    \label{fig:resolution}
\end{figure}

S\textsuperscript{4}G galaxies are $\sim \!$ 5 to 60 Mpc away from us.
To identify whether our analysis suffers any proximity bias, \Fig{resolution} plots intrinsic flattening $c/a$ of S\textsuperscript{4}G edge-on galaxies as a function of distance. 
Qualitative and quantitative inspections of \Fig{resolution} show that our intrinsic flattening values are independent of distance and thus free of resolution effects.
Therefore, distances do not bias our results through resolution effects.

Our study did not account for PSF effects which could introduce scattered light from the bulge or midplane into the disk's extended structure.
Such a treatment could employ the IRAC PSF of \citet{Hora2012}.
A symmetrised version of this PSF is required for such an effort, as all S\textsuperscript{4}G images are composed of two visits to each galaxy at different orientations by the \emph{Spitzer} Space Telescope.
An azimuthally averaged PSF would be needed to deconvolve our images. 

\begin{figure}
    \centering
    \includegraphics[width=\linewidth]{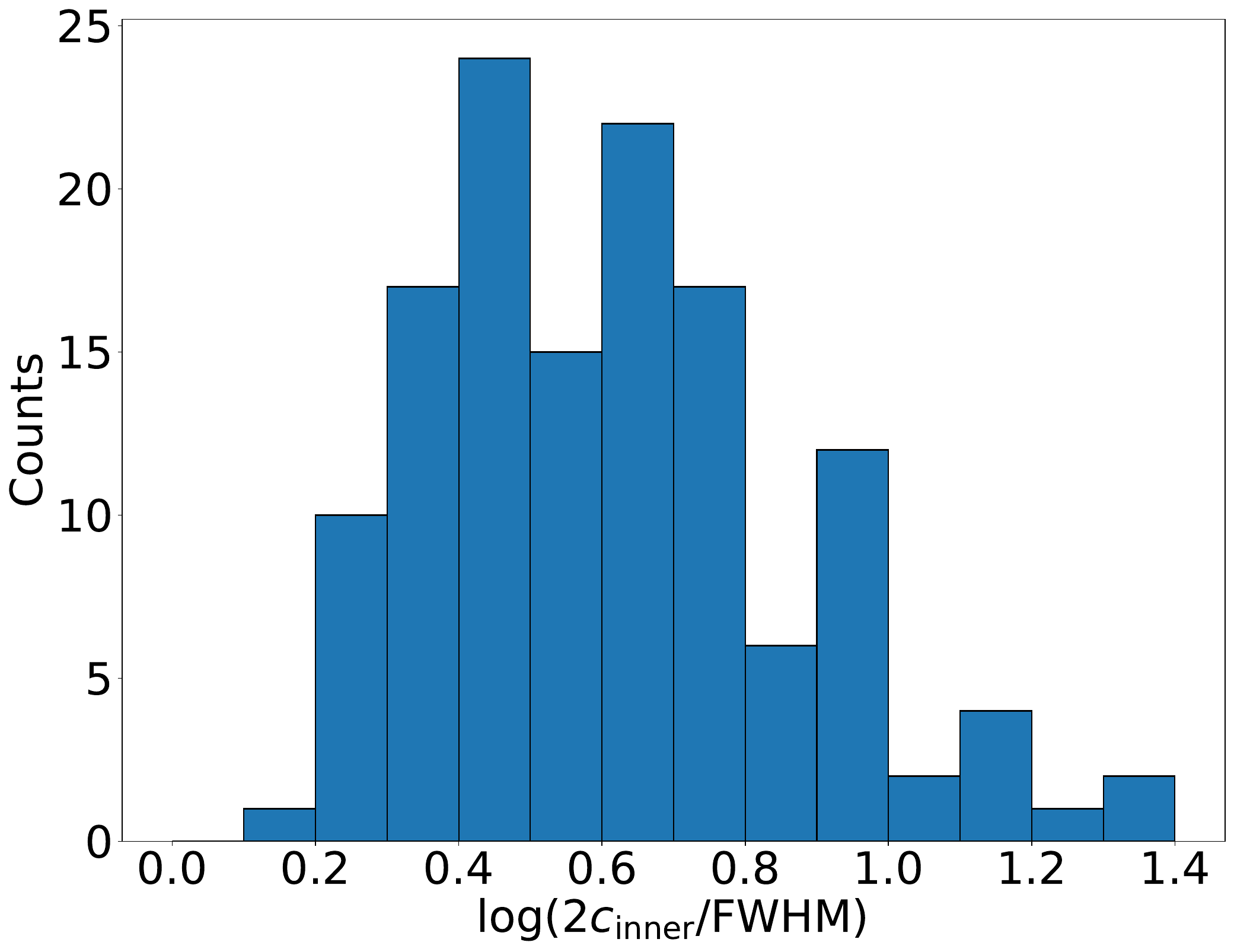}
    \caption{Histogram of the ratio $c$ from the inner cuts over the PSF FWHM.}
    \label{fig:c/PSF_hist}
\end{figure}

While the exploration of PSF effects is beyond the scope of the current paper, we can alleviate most concerns by comparing the scale of an approximate PSF to the shortest scale studied here.
We used a $2.1''$ FWHM Gaussian as an approximation for the S\textsuperscript{4}G PSF.
The shortest scale studied is the semi-minor axis of the inner cut.
These scales are compared by taking the ratio of $c$ to half the PSF FWHM, or $2c_{\rm{inner}}/${2.1\arcsec}, as shown in the log-binned histogram of \Fig{c/PSF_hist}.
For this ratio to be close to or less than one, the light scattered by the PSF would be of some concern. 
\Fig{c/PSF_hist} suggests that our results do not suffer significantly from PSF induced scattered light.
The mean value of the ratio $2c_{\rm{inner}}/${2.1\arcsec} is 4.9, with a smallest value of 1.6.
Still, a future analysis the distribution of $c/a$ ought to explore this question more closely.

\bibliography{bib}{}
\bibliographystyle{aasjournal}



\end{document}